\documentclass[journal]{IEEEtran}
\usepackage{color,soul}
\definecolor{darkcyan}{rgb}{0.0, 0.55, 0.55}
\soulregister\cite7
\soulregister\ref7
\soulregister\pageref7
\soulregister\gls7

\usepackage{glossaries}

\newglossaryentry{afe}{name={AFE},description={Auditory Front-end},
    first={\glsentrydesc{afe} (\glsentrytext{afe})}}

\newglossaryentry{amlttp}{name={AMLTTP},description={Auditory Machine Learning Training and Testing Pipeline},
    first={\glsentrydesc{amlttp} (\glsentrytext{amlttp})}}

\newglossaryentry{bic}{name={BIC},description={Bayesian information criterion},
    first={\glsentrydesc{bic} (\glsentrytext{bic})}}

\newglossaryentry{brir}{name={BRIR},description={binaural room impulse response},
    first={\glsentrydesc{brir} (\glsentrytext{brir})},
    plural={BRIRs},
    descriptionplural={binaural room impulse responses},
    firstplural={\glsentrydescplural{brir} (\glsentryplural{brir})}}

\newglossaryentry{bss}{name={BSS},description={Blind Source Separation},
    first={\glsentrydesc{bss} (\glsentrytext{bss})}}

\newglossaryentry{cnn}{name={CNN},description={Convolutional Neural Network},
    first={\glsentrydesc{cnn} (\glsentrytext{cnn})},
    plural={CNNs},
    descriptionplural={Convolutional Neural Networks},
    firstplural={\glsentrydescplural{cnn} (\glsentryplural{cnn})}}

\newglossaryentry{crnn}{name={CRNN},description={Convolutional Recurrent Neural Network},
    first={\glsentrydesc{crnn} (\glsentrytext{crnn})},
    plural={CRNNs},
    descriptionplural={Convolutional Recurrent Neural Networks},
    firstplural={\glsentrydescplural{crnn} (\glsentryplural{crnn})}}

\newglossaryentry{dnn}{name={DNN},description={Deep Neural Network},
    first={\glsentrydesc{dnn} (\glsentrytext{dnn})},
    plural={DNNs},
    descriptionplural={Deep Neural Networks},
    firstplural={\glsentrydescplural{dnn} (\glsentryplural{dnn})}}

\newglossaryentry{glm}{name={GLM},description={generalized linear model},
    first={\glsentrydesc{glm} (\glsentrytext{glm})},
    plural={GLMs},
    descriptionplural={generalized linear models},
    firstplural={\glsentrydescplural{glm} (\glsentryplural{glm})}}

\newglossaryentry{gmm}{name={GMM},description={Gaussian mixture model},
    first={\glsentrydesc{gmm} (\glsentrytext{gmm})},
    plural={GMMs},
    descriptionplural={Gaussian mixture models},
    firstplural={\glsentrydescplural{gmm} (\glsentryplural{gmm})}}

\newglossaryentry{hrir}{name={HRIR},description={head-related impulse response},
    first={\glsentrydesc{hrir} (\glsentrytext{hrir})},
    plural={HRIRs},
    descriptionplural={head-related impulse responses},
    firstplural={\glsentrydescplural{hrir} (\glsentryplural{hrir})}}

\newglossaryentry{ild}{name={ILD},description={interaural level-difference},
    first={\glsentrydesc{ild} (\glsentrytext{ild})},
    plural={ILDs},
    descriptionplural={interaural level-differences},
    firstplural={\glsentrydescplural{ild} (\glsentryplural{ild})}}

\newglossaryentry{itd}{name={ITD},description={interaural time-difference},
    first={\glsentrydesc{itd} (\glsentrytext{itd})},
    plural={ITDs},
    descriptionplural={interaural time-differences},
    firstplural={\glsentrydescplural{itd} (\glsentryplural{itd})}}

\newglossaryentry{mfcc}{name={MFCC},description={Mel-frequency cepstral coefficients},
    first={\glsentrydesc{mfcc} (\glsentrytext{mfcc})},
    plural={MFCCs},
    descriptionplural={Mel-frequency cepstral coefficients},
    firstplural={\glsentrydescplural{mfcc} (\glsentryplural{mfcc})}}

\newglossaryentry{rnn}{name={RNN},description={Recurrent Neural Network},
    first={\glsentrydesc{rnn} (\glsentrytext{rnn})},
    plural={RNNs},
    descriptionplural={Recurrent Neural Networks},
    firstplural={\glsentrydescplural{rnn} (\glsentryplural{rnn})}}

\newglossaryentry{snr}{name={SNR},description={signal-to-noise ratio},
    first={\glsentrydesc{snr} (\glsentrytext{snr})},
    plural={SNRs},
    descriptionplural={signal-to-noise ratios},
    firstplural={\glsentrydescplural{snr} (\glsentryplural{snr})}}

\newglossaryentry{sss}{name={SSS},description={Spatial Stream Segregation},
	first={\glsentrydesc{sss} (\glsentrytext{sss})}}

\newglossaryentry{tf}{name={TF},description={time-frequency},
	first={\glsentrydesc{tf} (\glsentrytext{tf})}}

\newglossaryentry{vm}{name={VM},description={von Mises},
	first={\glsentrydesc{vm} (\glsentrytext{vm})}}

\usepackage[noadjust]{cite}

\ifCLASSINFOpdf
  \usepackage[pdftex]{graphicx}
\else
\fi
\usepackage{amsmath}
\usepackage{amssymb}
\usepackage{mathtools}
\newcommand{\defeq}{\vcentcolon=}

\usepackage{booktabs}

\ifCLASSOPTIONcompsoc
  \usepackage[caption=false,font=normalsize,labelfont=sf,textfont=sf]{subfig}
\else
  \usepackage[caption=false,font=footnotesize]{subfig}
\fi
\usepackage{url}

\usepackage[capitalise]{cleveref}

\usepackage{siunitx}
\usepackage{framed}
{\begin{leftbar}\begin{quotation}}%
{\end{quotation}\end{leftbar}}

\begin{document}
%
\title{Joining Sound Event Detection and Localization Through Spatial Segregation}
%
%
%

\author{Ivo~Trowitzsch, Christopher~Schymura, Dorothea~Kolossa,
        and~Klaus~Obermayer
\thanks{The authors IT and KO are with the Neural Information Processing Group, Department of Electrical Engineering and Computer Science, Technische Universit{\"a}t Berlin. The authors CS and DK are with the Cognitive Signal Processing Group, Department of Electrical Engineering and Information Technology, Ruhr-Universit{\"a}t Bochum.}
}

%
%

\markboth{IEEE/ACM Transactions on Audio, Speech and Language Processing}%
{Shell \MakeLowercase{\textit{et al.}}: Bare Demo of IEEEtran.cls for IEEE Journals}
%

\IEEEpubid{This article has been accepted for publication in a future issue of this journal. Content may change prior to final publication. 2329-9290 \textcopyright 2019 IEEE.}


\maketitle

\begin{abstract}
Identification and localization of sounds are both integral parts of computational auditory scene analysis. Although each can be solved separately, the goal of forming coherent auditory objects and achieving a comprehensive spatial scene understanding suggests pursuing a joint solution of the two problems. This work presents an approach that robustly binds localization with the detection of sound events in a binaural robotic system. Both tasks are joined through the use of spatial stream segregation which produces probabilistic time-frequency masks for individual sources attributable to separate locations, enabling segregated sound event detection operating on these streams. We use simulations of a comprehensive suite of test scenes with multiple co-occurring sound sources, and propose performance measures for systematic investigation of the impact of scene complexity on this segregated detection of sound types. Analyzing the effect of spatial scene arrangement, we show how a robot could facilitate high performance through optimal head rotation. 
Furthermore, we investigate the performance of segregated detection given possible localization error as well as error in the estimation of number of active sources. Our analysis demonstrates that the proposed approach is an effective method to obtain joint sound event location and type information under a wide range of conditions.

\end{abstract}


%
\IEEEpeerreviewmaketitle


\section{Introduction}

%
%
%
%
\IEEEPARstart{R}{ealistic} 
aural environments consist of numerous co-occurring different sounds emitted from sources distributed in space. Computational auditory scene analysis thus involves the development of models that draw information from audio streams and assign semantic labels to auditory objects. For instance, a robotic system that is specialized to search and rescue missions should be able to detect the presence of a fire, an alarm that is going off, screaming victims, or a crying baby, and localize them. Two key issues therefore are (a) detecting sound events and their types within that stream, commonly called audio or sound event detection (\emph{SED}), and (b) localizing the corresponding sources emitting the sounds, denoted sound source localization (\emph{SSL}).  This work investigates the combination of the two: joint sound event localization and detection (\emph{SELD}). 

For comprehensive understanding of acoustic (or any) scenes, it is not only necessary to know what is there and where there is something, but instead to know \emph{what is where}. However, coherently attributing ``type'' and ``location'' to auditory objects in scenes with multiple simultaneously active sources is notably more difficult than performing SED or SSL individually, which is why there are only few works on the topic so far (\cite{Lopatka2016,butko2011two,Ma2018,Chakraborty2014,Grobler2017,May2012binaural,hirvonen2015classification,He2018,adavanne2018sound}).

\IEEEpubidadjcol

There are four different fundamental approaches to joining sound event detection and source localization:

\subsubsection{Temporal correlation}
Associating type and location of sounds through temporal correlation. This however is not possible for multiple sounds starting at the same time; and difficult for moving sources. If sources move (temporarily) to the same location, tracking gets lost. 
 
\subsubsection{Sound-type masked SSL}
Attending to streams related to individual sound events. This ``focus'' can be created through masking the input such that a particular sound known to be active is ``passed through'' to SSL, and other sounds or noise are suppressed. Such masking is feasible in time-frequency domain if sound events exhibit specific frequency signatures. The subsequent localization then produces locations associable to these events. However, not all sound event classes exhibit coherent (and narrow) frequency signatures -- for instance ``alarm'' is more of a semantic class, and can range from electronic beeps to fire bells; or ``piano'' ranges from very low to high frequencies. Also, the approach is likely to fail for co-occurring sound events with similar frequency patterns. 

A work following this approach is presented in \cite{Ma2018}. While the system is advertised and analyzed with regard to improving SSL, not SELD, it effectively joins sound event type and sound source location information.

\subsubsection{Spatially masked SED}
Attending to streams related to individual source locations. This implies masking of the input such that only sound from a particular direction is passed through to SED, and sound from other directions is suppressed. Such masking is technically doable in time domain through beam-forming, or in time-frequency domain through spatial segregation, attributing individual time-frequency-bins to particular directions. Sound events detected on the spatial streams are then associated with a location. Efficient masking depends on spatial separation of sources and hence dissimilar spatial cues. 

Following approach \emph{3}, there is one work using beam-forming \cite{Grobler2017}, and one comprehensive work \cite{May2012binaural} in which spatial masks are applied to the time-frequency feature space, on which a \gls{gmm} detects speech. \cite{Chakraborty2014} mix elements of approach \emph{3} (beam-forming) and \emph{2} (using prior sound event detection information on the full stream).

\begin{figure*}[t]
    \centering
    \includegraphics[width=0.99\textwidth]{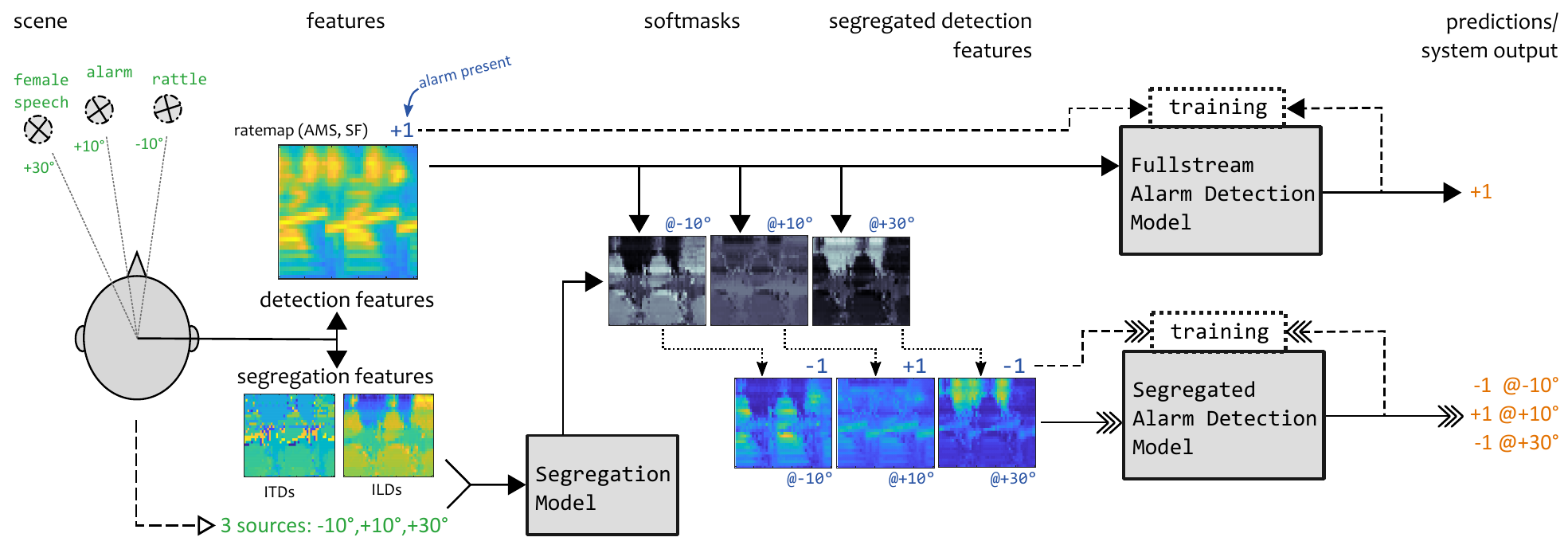}
    \caption{From binaural scenes to localized detections. Exemplary scene with three sources, at \ang{-10} (emitting female speech), \ang{+10} (alarm sound), and \ang{+30} (rattling sound). From ear signals, detection and segregation features are computed and cut into blocks of \SI{500}{\milli\second} (amplitude modulation spectrograms (AMS) and spectral features (SF) omitted for clarity). Together with input about number of active sources and their azimuths (ground truth at training time, systematically perturbed or ground truth at testing time for our analysis; estimated or set values in a deployment system), the segregation model produces one softmask for each spatial stream, that is, for each azimuth. Each softmask gets applied to the detection features, such that one set of features is formed per stream. Labels about the presence of target sound events (alarm, in this case) are attached according to the associated azimuth in training. Segregated features are then input to the segregated detection models, which are trained to predict the presence of target events in the passed blocks (in the depicted example, all predictions are correct). Fullstream models -- not part of the segregated detection but complementary and for comparison -- get non-masked detection features and detect the presence of target events in the full mixture.  }
    \label{fig:informationFlowDiagram}
\end{figure*}

\subsubsection{Joint SELD}
Building models that by construction detect localized sound events. Such models do not localize and detect separately or subsequently (as in approaches 2 and 3), but instead produce joint attributes from the start. While -- because of the implicit combination of approaches 1-3 -- this approach should in principle be the most powerful, it also requires the most powerful model, more difficult to train and to understand.

Three more recent publications present fully joint SELD systems following approach \emph{4}, all employing \glspl{dnn}: \cite{hirvonen2015classification} and \cite{He2018} build speech detection and localization models, demonstrating the feasibility of training joint sound event detection and localization models based on \glspl{dnn} \cite{hirvonen2015classification} and \glspl{cnn} \cite{He2018}, although the former only with one source active at a time, and the latter with results difficult to judge, being restricted to overall averages. The so far most comprehensive paper on the topic \cite{adavanne2018sound} features a fully joint SELD system based on a \gls{crnn} model.



\emph{In the present work}, we are following approach \emph{3}, detecting sound events on spatially segregated streams. Apart from the immediately related works mentioned above, this approach is related to \gls{bss} techniques from the field of digital signal processing~\cite{Park2005, Do2011, Maazaoui2012, Nakanishi2015, Faller2017}. The employed spatial segregation model, computing softmasks in time-frequency space as similarly described in~\cite{kolossa2004nonlinear,Harishkumar2014, Ma2018}, serves as a processing step for associating auditory features later used for sound event detection with specific sound source locations. 

Compared to approach \emph{2}, sound event detection on spatial streams has the advantage of enabling localized identification of multiple sources of the same type or with similar frequency ranges active. Compared to approach \emph{4}, this approach is feasible also with less powerful models classes, faster to train, and easier to understand. Furthermore, systems following approach \emph{3} are modular, which enables work and research on the individual components. 

We use the spatially masked auditory features for sound event detection in a scheme called multi-conditional training, presented in \cite{Trowitzsch2017Robust} for robust binaural SED modeling. Although \glspl{dnn} in various forms are predominant in sound event detection by now (\cite{Hertel2016,Phan2016,Cakir2017,Jeong2017,Hayashi2017,Li2017,DCASE2016workshop,DCASE2017Workshop,DCASE2018Workshop}), we chose to stick to the LASSO-models used in \cite{Trowitzsch2017Robust}, which are very easy and fast to train and test, and have shown to produce decent performance with the employed ratemap and amplitude modulation spectrogram features in above reference. We opted to rather perform and provide extensive tests and qualitative analysis of the system instead of demonstrating the best performance possible. 

Contrary to the related speech SELD system in \cite{May2012binaural}, we decided to train the localized SED models with mask application \emph{included} and multi-conditionally with overlapping sources, rather than training on clean data and applying masks during testing together with a missing data approach. Results of \cite{May2012binaural} show a strong dependence on \gls{snr} even in regimes with the noise exhibiting less energy than the target signal, and we believe that our data-driven approach potentially leads to more robust identification.

\begin{figure*}[t]
    \centering
    \subfloat[Bisected]{\includegraphics[width=0.25\textwidth]{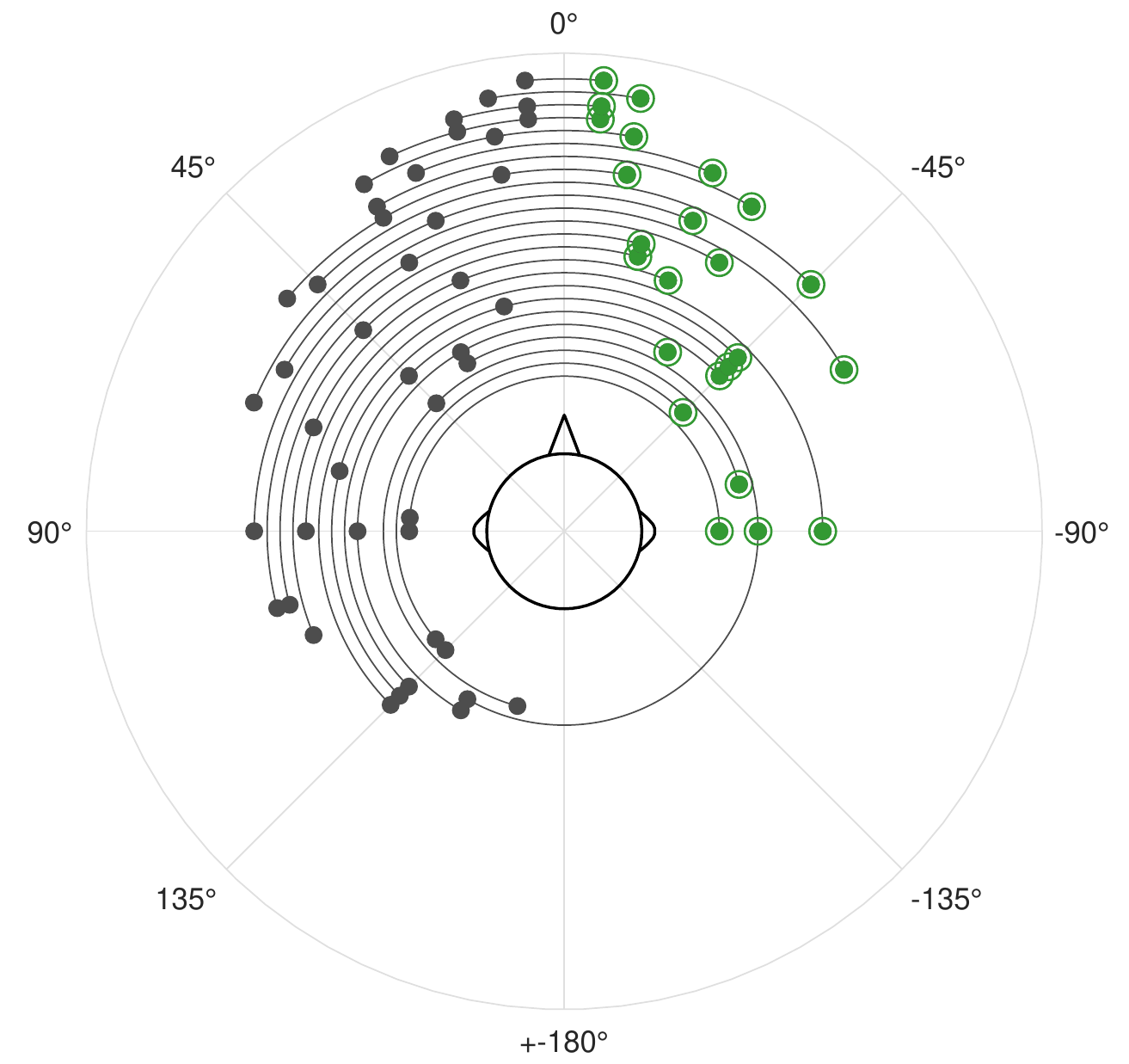}%
        \label{fig:sceneconfig_1}}
    \hfil
    \subfloat[Target@0$^{\circ}$]{\includegraphics[width=0.25\textwidth]{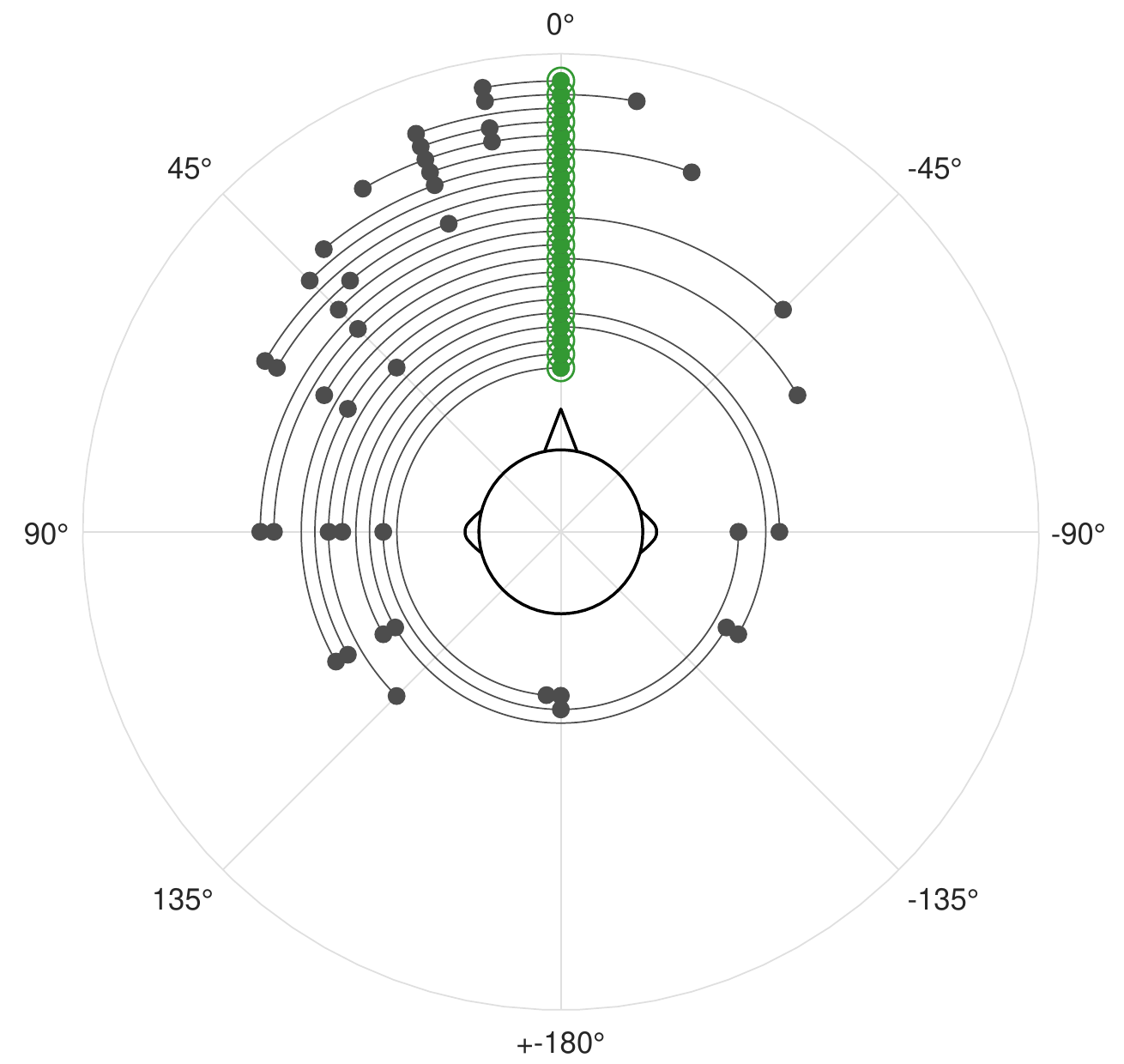}%
        \label{fig:sceneconfig_2}}
    \hfil
    \subfloat[Front-left]{\includegraphics[width=0.25\textwidth]{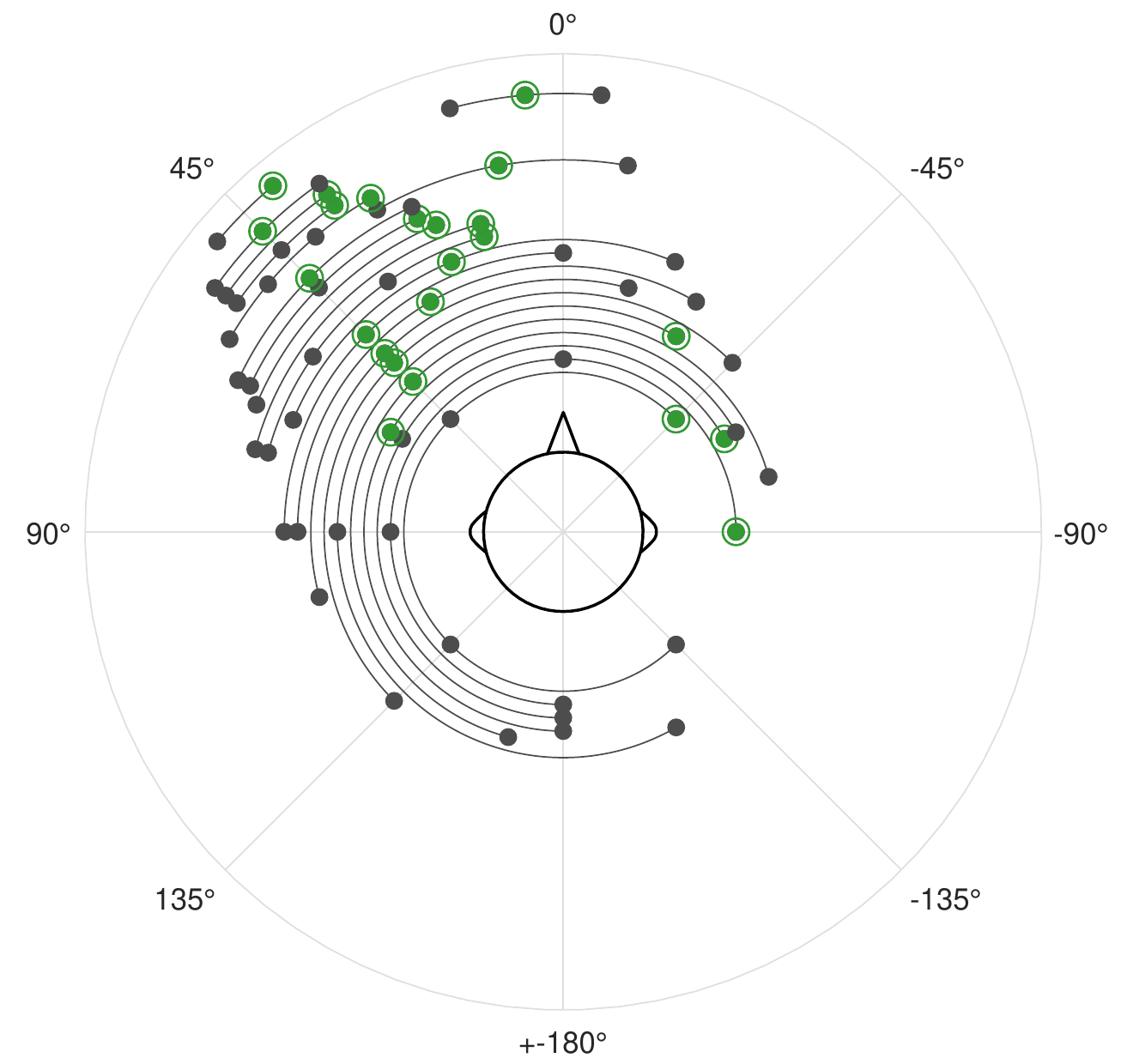}%
        \label{fig:sceneconfig_3}}
    \hfil
    \subfloat[Ear-centered single-hemisphere]{\includegraphics[width=0.25\textwidth]{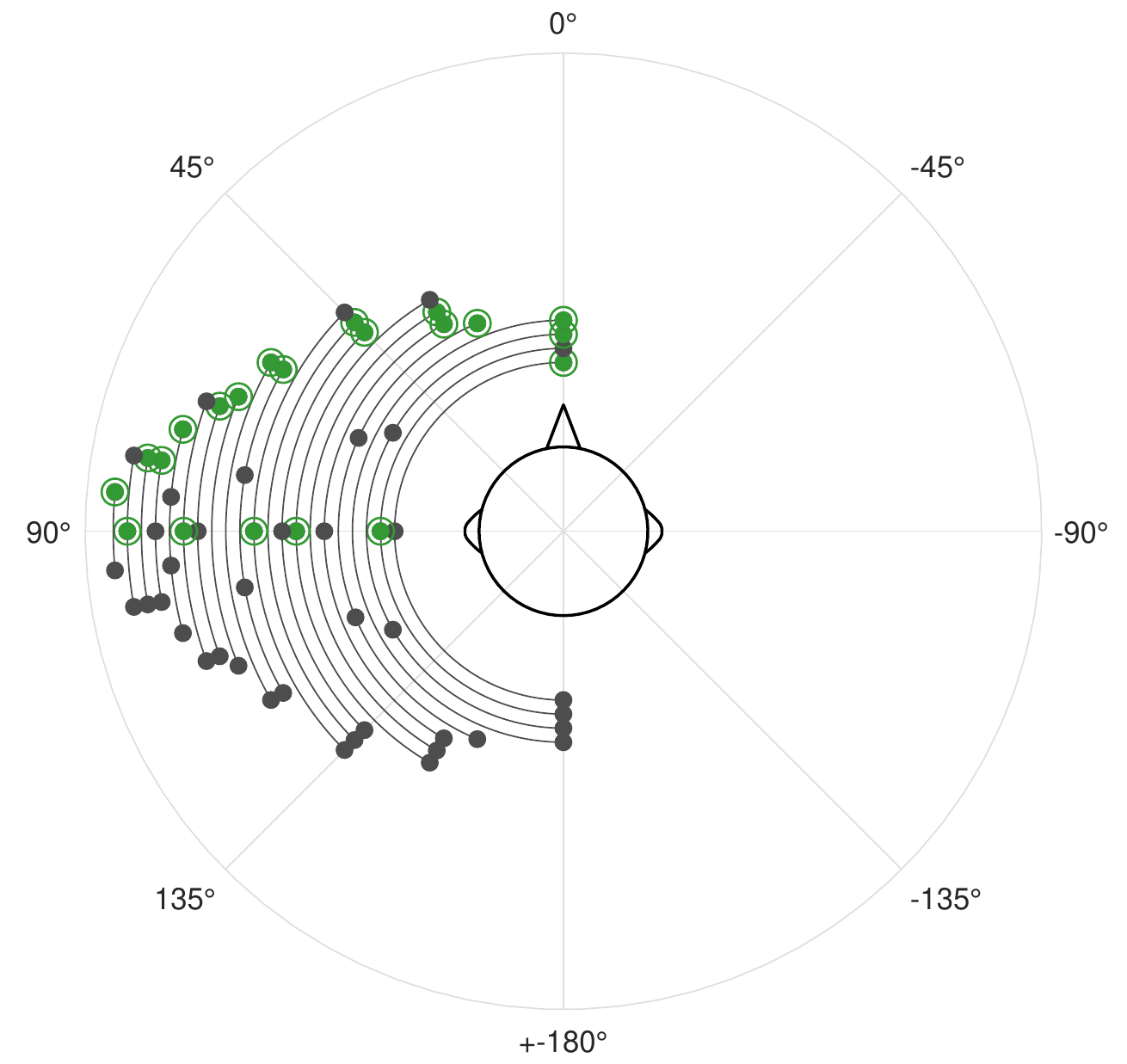}%
        \label{fig:sceneconfig_4}}
    \caption{Test scene configurations (only scenes with at least two sources), sorted into the different \emph{scene modes}. Black filled circles depict distractor sources, target sources (green) are highlight by an enclosing open circle. Each scene is indicated by one circle fragment. The head is at the center.}
    \label{fig:sceneconfigs}
\end{figure*}

This work was started in the scope of the {\scshape Two!Ears} Project (\url{http://twoears.eu}, \cite{twoears}), which aimed for comparability to human performance and whose goal was to enable better computational understanding of auditory and multi-modal perception. Hence, we restrict ourselves to \emph{binaural} processing, simulating ear signals of a humanoid robot. The only related articles covering binaural analysis are \cite{Ma2018, May2012binaural}. Most commonly arrays with more than two microphones are used; in general spatial segregation and localization often rely on the availability of multiple microphones for maximum performance \cite{Nadiri2014,Gannot2017,He2018,adavanne2018sound}. 

We propose a method of how to construct a system that is able to produce joint sound event type and location information \emph{given} sound source locations and number of sources; we do not suggest how to estimate the number of sources and their locations. This permitted greater focus on the segregation part and systematic evaluation with respect to quality of these inputs.

We provide a comprehensive analysis of the resulting SELD system with regard to acoustic conditions --- particularly the influence of spatial source distribution has not been studied before in this context --- and dependency on correct input of number of active sources and quality of localization. 
Finally, the effects of diffuse noise and reverberation are discussed.

For better comparison of detection performances, \emph{fullstream} SED models operating on the full mixture (in contrast to the segregated streams) have been trained and tested on exactly the same multi-conditional auditory data. Fullstream and segregated detection models share all methodology and operating principles except for, of course, the spatial segregation and localization of segregated streams that is added in segregated detection. These models are no obligatory components of our proposed system; but could be complementary, as will be seen later.

Our system as presented here is a proposition of how to join sound event detection and localization, and how to analyze and measure performance of such a system. The system and test scenes, together with the suggested performance measures, can serve as benchmarks for other SELD systems or components. \cref{fig:informationFlowDiagram} depicts the system and its information flow.


\section{Methods}


\subsection{Sound Data}

The NIGENS database \cite{trowitzsch2019nigens} was used to create the acoustic scenes for training and testing. This database provides audio files of $13$ different event classes: alarm, crying baby, crash, barking dog, running engine, burning fire, footsteps, female speech, male speech, knocking, phone, piano, screams (we combined male and female screams into one class), and a ``general'' class with sounds of all kinds, exhibiting as much variety as possible. 
All sound events included are isolated without superposition of ambient or other foreground sources. The sound files are annotated with on- and offsets of the included sound events, many files include several event instances. The ``general'' sounds only served as negative examples for the classifiers and as distractor signals (see \cref{sec:scenes,sec:methods_training}).

Containing $1017$ wav files, NIGENS is the currently largest database of truly isolated sound events annotated with event on- and offset times.


\subsection{Binaural Auditory Scenes}\label{sec:scenes}


A set of binaural auditory scenes was rendered for training the detection models, and another set for testing. These scenes consist of different numbers of sound sources (one to four) at different locations --- whenever the term location is used throughout this work, it refers to the azimuth relative to the head, disregarding distance and elevation. 

To create the two-channel ``ear signals'', we used the binaural simulator of the \textsc{Two!Ears} system \cite{two_ears_team_2018_1458420}. The simulator convolved the audio source with an anechoic \gls{hrir} measured with a Knowles Electronic Manikin for Acoustic Research (KEMAR) head, resulting in two-channel ``ear signals'' \cite{Wierstorf2011afree}.

Binaural simulation was conducted for each source separately to allow control over the energy ratio of the sources \emph{in the ear signals}. To this end, the resulting ear-signal streams for each source were mixed at defined ratios of squared amplitudes averaged over both binaural channels and time, only including times of sound activity to not influence the ratio by periods of silence.  
We later refer to these ratios as \glspl{snr} even though there is no ``classic'' noise involved.
The \gls{snr} was fixed such that it was the \gls{snr} of the ``target'' to each individual ``distractor'' source. Distractor sources never simultaneously emitted a sound from the same class as the sound emitted by the target source. This way we were able to control event-wise \glspl{snr} and evaluate systematically.
    
\emph{Eighty training scenes} were defined for multi-conditional training, as introduced in \cite{Trowitzsch2017Robust}. However, due to the increased number of free parameters of scenes with more than two sources, it seemed more efficient to randomly sample the parameter space compared to manual definition of scenes. We randomly chose 
\begin{itemize}
    \item the number of sources (one to four)
    \item the azimuths between head and sources (uniformly between \ang{+-180}, discretized to \ang{22.5}-steps\footnote{This discretization serves computation efficiency; \ang{22.5} is a compromise between smaller number of renderings and more dense spatial sampling.})
    \item the \glspl{snr} between target and other sources (uniformly between \SI{-20}{\decibel} and \SI{+20}{\decibel}).
\end{itemize}

For \emph{testing, 468 scenes} were defined such that it would be possible to look at only one scene parameter changing and keeping the others constant --- the high number of scenes compared to the training set is due to this constraint. The following parameters were varied:
\begin{itemize}
    \item the number of sources (one to four)
    \item the \glspl{snr} between target and distractor sources (\SI{-20}{\decibel}, \SI{-10}{\decibel}, \SI{0}{\decibel}, \SI{+10}{\decibel}, \SI{+20}{\decibel})
    \item the azimuth difference between sources (\ang{0}, \ang{10}, \ang{20}, \ang{45}, \ang{60}, \ang{90}, \ang{120}, \ang{180})
    \item the ``scene mode'' (depicted in \cref{fig:sceneconfigs}): 
        \begin{enumerate}
            \item \emph{bisecting}: the nose (\ang{0} azimuth) points between target and distractor source(s)
            \item \emph{target@0}: the nose points towards the target source
            \item \emph{front-left}: sources are mainly between \ang{0} and \ang{90}; they are not bisected and targets are not at \ang{0}, and they are not symmetric around the ear
            \item \emph{ear-centered}: sources are distributed in the left hemisphere symmetrically around the ear (\ang{90})
        \end{enumerate}
    \item the position of the target among the sources: either at one end, or (only for three-source scenes\footnote{Only for three-source scenes to save computation time.}) at the center.
\end{itemize}
No differentiation was made between left and right, since head and scenes are symmetric.

For all scenes, each of the NIGENS files was used once as target source sound. Sounds shorter than $30s$ were looped.

Exact definitions of training and test scenes can be found in the supplementaries, as well as the code to replicate them together with AMLTTP \cite{amlttp30} and NIGENS \cite{trowitzsch2019nigens}.


\subsection{Spatial Stream Segregation}\label{sec:segregation} 

To produce spatially segregated streams which subsequently can be analyzed by sound event detectors, the segregation model computes softmasks in the time-frequency-space for a set of specified sound source locations, given the actual spatial cues (\glspl{itd}\footnote{Time-frequency-binned \glspl{itd} are determined as the lags of the most prominent peaks on the normalized cross-correlation functions computed for short time frames on the gammatone-filtered inner-haircell representations of the ear signals.}
and \glspl{ild}). That is, for each of these locations, the segregation model computes a value between $0$ and $1$ for every time-frequency-bin, corresponding to the likelihoods of their spatial cues having been produced by a source at this location.
The general information flow concerning the stream segregation stage and its embedding into the segregated detection system is depicted in \cref{fig:informationFlowDiagram}. Blocks of \glspl{itd} and \glspl{ild} in time-frequency-representation, as well as the estimated number of active sources with corresponding locations serve as inputs to the segregation model. 

At the core of the spatial stream segregation, \glspl{glm}~\cite{Dobson2002} are used as mapping functions from azimuthal source location to prototypical binaural observations \(\boldsymbol{y}_{kl}~=~\begin{bmatrix}\tau_{kl} & \delta_{kl}\end{bmatrix}^{T}\) with
\gls{itd} \(\tau_{kl}\) and 
\gls{ild} \(\delta_{kl}\) at each time frame \(k\) and frequency channel \(l\). The underlying observation models are represented as
\begin{equation}
\label{eqn:obs_model}
\boldsymbol{y}_{kl} = \boldsymbol{g}_{l}(\phi) + \boldsymbol{n}_{kl},
\end{equation}
where \(\boldsymbol{g}_{l}(\phi)\) is a mapping function of an azimuth angle \(\phi\) and \(\boldsymbol{n}_{kl} \sim \mathcal N(\boldsymbol{0},\,\boldsymbol{R}_{l})\) is an additive Gaussian noise term with frequency-dependent covariance matrix \(\boldsymbol{R}_{l}\). The mapping function is realized as a \gls{glm}
\begin{equation}
\label{eqn:obs_reg_model}
\boldsymbol{g}_{l}(\phi) = \begin{bmatrix}
\beta_{l0}^{\tau} + \sum_{n=1}^{N} \beta_{ln}^{\tau} \sin (n \cdot \phi) \\
\phantom{.} \\
\beta_{l0}^{\delta} + \sum_{n=1}^{N} \beta_{ln}^{\delta} \sin (n \cdot \phi)
\end{bmatrix},
\end{equation}
which is based on trigonometric functions to account for the circular nature of the azimuthal source positions. Herein, \(N\) is the maximum order of the regression function and \(\beta_{ln}^{\{\tau,\delta\}}\) represent the regression coefficients, which are estimated via linear regression using anechoic \glspl{hrir}~\cite{Wierstorf2011afree} with white noise as stimulus signals. The residuals obtained after training are used to estimate the noise covariance matrix~\(\boldsymbol{R}_{l}\). The best-fitting model order \(N\) was determined using a selection process based on the \gls{bic}~\cite{Nielsen2013}. 
The model introduced in Eq.~\eqref{eqn:obs_reg_model} essentially performs a fitting of \glspl{itd} and \glspl{ild} via sinusoidal functions, which allows to derive a continuous representation of these cues.

Given a set of \(M\) estimated azimuthal source locations \(\{\phi_{i}\}_{i = 1}^{M}\) and actual binaural observations \(\tilde{\boldsymbol{y}}_{kl}\), the model described in \cref{eqn:obs_model,eqn:obs_reg_model} produces a softmask weighting factor for the \(i\)-th source at time-step \(k\) and frequency-channel \(l\) according to 
\begin{equation}
\label{eqn:soft_mask}
m_{kl}^{(i)} = \frac{p(\tilde{\boldsymbol{y}}_{kl}\,|\,\boldsymbol{g}(\phi_{i}),\boldsymbol{R}_{l})}{\sum_{j=1}^{M} p(\tilde{\boldsymbol{y}}_{kl}\,|\,\boldsymbol{g}(\phi_{j}),\boldsymbol{R}_{l})},
\end{equation}
where \(p(\tilde{\boldsymbol{y}}_{kl}\,|\,\boldsymbol{g}(\phi_{i}),\,\boldsymbol{R}_{l})\) is the likelihood (given the observation model \(\boldsymbol{g}(\phi)\)) that the actual binaural cues \(\tilde{\boldsymbol{y}}_{kl}\) were produced by a source at the \(i\)-th azimuth at time-step \(k\) and frequency-channel \(l\), 
\begin{equation}
\label{eqn:likelihood}
p(\tilde{\boldsymbol{y}}\,|\,\boldsymbol{g}(\phi),\boldsymbol{R}) = \frac{\exp(-\frac{1}{2}(\tilde{\boldsymbol{y}}-\boldsymbol{g}(\phi))\boldsymbol{R}^{-1}(\tilde{\boldsymbol{y}}-\boldsymbol{g}(\phi))')}{\sqrt{|\boldsymbol{R}|(2\pi)^2}},
\end{equation}
and \(m_{kl}^{(i)}\) corresponds to the probability that the \(i\)-th source is dominant at time-step \(k\) and frequency-channel \(l\). 
An example of softmasks produced by Eq.~\eqref{eqn:soft_mask} is depicted in \cref{fig:informationFlowDiagram}.

It should be noted that the proposed spatial segregation model yields a rather general approximation to the required binaural cues. Even though training is conducted on anechoic \glspl{hrir} in this work, specific adaptations using, e.g., individualized \glspl{hrir} or \glspl{brir} are generally possible and can be applied according to the corresponding acoustic scenario.


\subsection{Fullstream Detection Model Input}\label{sec:methods_sd_features} 

The simulated binaural auditory scenes were processed by the auditory front-end of the \textsc{Two!Ears} system \cite{two_ears_team_2018_1458420} to obtain the following representations:
\begin{itemize}
    \item \emph{ratemaps}: spectrograms resembling auditory nerve firing rates. Ratemaps are computed by smoothing the gammatone-filtered (32 channel) inner-hair-cell signal with a leaky integrator and binning into overlapping frames of \SI{20}{\milli\second} length (\SI{10}{\milli\second} shift). \cite{Glasberg1990derivation,Patterson1996functional,Cooke200126robust,May2012binaural}
    \item \emph{spectral features}: 14 different features summarizing the spectral content of the ratemap for each time frame: Centroid, Spread, Brightness, High-frequency content, Crest, Decrease, Entropy, Flatness, Irregularity, Kurtosis, Skewness, Roll-off, Flux, Variation. \cite{Peeters2011timbre,Tzanetakis2002musical,Jensen2003real,Misra2004spectral,Lerch2012introduction, Geiger2013Large,Marchi2016Up}
    \item \emph{amplitude modulation spectrogram}: Each channel of the inner-hair-cell representation is  analyzed by a bank of logarithmically scaled modulation filters. We used \emph{16 frequency channels} and \emph{8 modulation filters}. \cite{Moritz2011amplitude,May2014computational}
\end{itemize}
For all representations, gammatone center frequencies ranged from $80$~Hz to $8$~kHz linearly spaced on the ERB scale. 
    
All representations were split into overlapping blocks of \SI{500}{\milli\second} (with a shift length of \SI{333}{\milli\second}). 
For each block, a feature vector was constructed as input to the sound type classification: first, representations were averaged over left and right channels\footnote{In \cite{Trowitzsch2017Robust}, channel-average features are compared with two-channel features.}, and the first two discrete time derivatives were computed. Features were then aggregated from L-statistics\footnote{L-statistics\cite{Hosking1990moments} are shown to be more robust than conventional statistics, particularly for higher moments and little data \cite[Ch.~9]{David2003Order}.} (L-mean, L-scale, L-skewness, L-kurtosis) computed over time. This amounted to $1,091$ dimensions per feature vector\footnote{For details of the feature construction see supplementary informations.}.

Sound event onset and offset times were used to label each block (and thus corresponding feature vector) according to whether the target class was present (+1) or absent (-1) within the block. A sound event was defined present if either it occupied at least \SI{75}{\percent} of a block, or (for short events), if at least \SI{75}{\percent} of the sound event was included in the block. Blocks with occupation of less than these \SI{75}{\percent}, but more than \SI{0}{\percent}, were excluded from training and testing because we considered them ambiguous.


\subsection{Segregated Detection Model Input}\label{sec:methods_segId} 

Segregating into spatial streams takes place after the generation of the different auditory representations and their segmentation into blocks (see \cref{sec:methods_sd_features}), and before the construction of feature vectors from the blocked auditory representations. 

The segregation model produces a set of probabilistic time-frequency softmasks, with the number of masks corresponding to the number of \emph{active} sources in a block\footnote{Actually of the number of locations with active sources (active spatial streams) --- two sources at the same location have to count as one.}. Sources with (mean) block energy above \SI{-40}{\decibel} of their maximum energy over the whole scene instance were defined active (in this block).

These masks were applied (through multiplication) to the ratemaps and amplitude modulation spectrograms; spectral features were then computed from the masked ratemaps. Hence, one set of masked representations per spatial stream was produced, such that one feature vector per spatial stream could be generated. \cref{fig:informationFlowDiagram} summarizes the data processing steps.

Since each mask is generated based on a presumed location of active sound source(s), each mask is associated with this particular location. Ergo, each feature vector for the detection model is attributable to this location --- and hence each detection itself, which is why we also call the output of the segregated detection models also localized detection.
    
Both for training the detection models and testing their performance, labels indicating the presence or absence of target sound types are needed. If a block was labeled negative (see \cref{sec:methods_sd_features}) before segregation, all segregated blocks were labeled negative. If a block was labeled positive before segregation, the segregated feature vector associated to the location \emph{closest} to the target source was labeled positive, and the others negative.


\subsection{Model Training}\label{sec:methods_training} 

Two types of models were trained: \emph{fullstream} detection models, operating on the full (mixed) stream, features, and labels as described in \cref{sec:methods_sd_features}, and \emph{segregated detection} models, operating on the segregated streams, features, and labels as described in \cref{sec:methods_segId}. Apart from this difference in input and a difference in sample\footnote{The combination of a feature vector and the respective label is a \emph{sample}.} weighting (elaborated on below), both model types were trained identically, as described in this section.

In \cite{Trowitzsch2017Robust}, the impact of superimposed distracting sources was systematically investigated, and demonstrated how robust models are obtained by including a range of conditions in the training data, a procedure called \emph{multi-conditional training}. We used multi-conditional training over all 80 scenes for all our models in this work, extending it to an even wider range of included conditions (see \cref{sec:scenes}).

For all sound types, binary classifiers were trained in an one-vs-all scheme. As classifying model, the ``Least Absolute Shrinkage and Selection Operator'' (LASSO) \cite{Tibshirani1996regression} (utilizing the ``GLMNET'' package \cite{Friedman2010regularization,Qian2013glmnet}) was used, a linear logistic regression model with an $L_1$ penalty for the regression coefficients. This penalty leads to sparse models by forcing many regression coefficients to zero, making LASSO a classification method with an embedded feature selection procedure. 
For adjusting the regularization parameter $\lambda$ (determining the strength of the $L_1$ regularization term) value, six-fold stratified cross-validation on the training set was performed. The value with the best cross-validation performance was chosen and used to train the model on the full training set. 

200k samples were used to actually train each model (due to memory restrictions of GLMNET)\footnote{As the LASSO model has few free parameters (number of features + 1), this amount was enough. Actually performance saturated around 120k samples.}, subsampled from the complete training set. The sub-sampling process enforced using equally many samples from each sound file's mixture, so that long sound files would not be overrepresented in the training set. Samples were weighted in training such that the total sample weight of all samples from scenes with 1, 2, 3 and 4 sources, respectively, was equal.

All scene generation, data processing, model training and model testing was done using the open-source \emph{Auditory Machine Learning Training and Testing Pipeline} \cite{amlttp30}, which wraps all steps described in \crefrange{sec:scenes}{sec:evaluation}.


\subsection{Model Testing}\label{sec:methods_testing}

Data were split into a training set for model building and a test set for estimating the generalization performance of the classifiers. To ensure that a block from the training set and a block from the test set never contained parts of the same sound file, training-test and cross-validation splits were conducted at the level of sound files. The set of sound files for each class (including the general class) was split into training set (\SI{75}{\percent}) and test set (\SI{25}{\percent}). Only the sounds from the training set were used to generate the auditory scenes for building the classification models, and only the sounds from the test set were used to generate the scenes for evaluating the prediction performance. The chosen ratio of training and test set balances between necessary training data to generate good models, and desirable predictive power of the test data. For both, the number of original sound files per class is the more essential parameter compared to the number of produced scenes from them.
    
While training was conducted multi-conditionally, tests were performed on individual scenes in order to conclude on relations between scene parameters and performances.

Blocks in which not all sources were \emph{active} (since sources emit sounds that also exhibit silences), got removed to better reflect the influence of the number of sources. All remaining samples from the test set were used without further sub-sampling, amounting to about 12 million samples tested with each fullstream model and about 30 million with each segregated detection model.


\subsection{Segregation Model Input Perturbation}\label{sec:methods_perturbation}

As described in \cref{sec:segregation,sec:methods_segId}, the segregation model needs input on the number of active spatial streams and on their azimuths. For training, ground truth knowledge was used. For testing, we implemented three different modes:
\begin{enumerate}
    \item Using ground truth for both data.
    \item Using ground truth for the number of active streams, but perturbing the location information of those streams. This perturbation is conducted by adding random azimuth values drawn from a normal distribution with sigma of \ang{5}, \ang{10}, \ang{20}, \ang{45}, and \ang{1000} to each block's location. The latter basically corresponds to drawing locations uniformly. Note that we did \emph{not} change the sources' locations in the scenes, but \emph{only the information about them} given as input to the segregation model, and hence the azimuth associated with a block.
    \item Using ground truth for the locations of active streams, but perturbing the data about the active source number. A uniformly drawn random number between -2 and +2 was added to the number of streams ground truth (thresholding downwards at 1). In case of a reduction, the respective number of locations handed to the segregation model was removed randomly. In case of an increase, locations drawn randomly from a uniform distribution between \ang{0} and \ang{360} were added to the segregation model input.
\end{enumerate}

\subsection{Performance Measurement and Evaluation}\label{sec:evaluation}

Training and testing performance was measured utilizing balanced accuracy (BAC). As argued in \cite{Trowitzsch2017Robust}, BAC is preferable over F-score and error rate because of their dependence on the data distribution --- it is closer to an interpretation of ``informedness'' of a classifier. The same line of argumentation can be found in \cite{POWERS2011} promoting Bookmaker Informedness (also known as Youden's J Statistic \cite{youden1950index}), which is a scaled equivalent of BAC.

For segregated detection training, BAC had to be adjusted: there are many negative segregated samples originating from fullstream blocks without positive present --- in the following sub-indexed ${npp}$ ---, however, there also exist segregated \emph{negative} samples originating from fullstream blocks with a \emph{positive present in another stream}, in the following sub-indexed ${pp}$. These, however, have a much lower proportion than the ${npp}$ negatives and would, if not up-weighted, have minor influence on training. This would result in worse localized detection performance, because of too low cost of not discriminating between target and distractor streams. Thus, we defined $BAC_{sw}$ ($sw$ for stream-wise) for segregated detection:
\begin{align}\label{eq:bac_sw}
BAC_{sw}& \defeq 0.5 \cdot SENS + 0.5 \cdot SPEC_{sw},\text{ with }\\
SPEC_{sw}& \defeq 0.5 \cdot SPEC_{pp} + 0.5 \cdot SPEC_{npp},\notag \\
SENS& \defeq TP / (TP+FN),\notag \\
SPEC& \defeq TN / (TN+FP).\notag
\end{align} 

While $BAC_{sw}$ summarizes performance in one number so that the models can be optimized, it is difficult to gain insight into the actual behavior of the models through that number. Two different aspects of segregated detection performance are interesting: time-wise detection performance, and localized detection performance.

\begin{table}[t]
    \caption{Measures \& nomenclature overview}
    \label{tab:measures}
    \centering
    \begin{tabular}{p{0.15\linewidth}p{0.75\linewidth}}
        \toprule
        \multicolumn{2}{p{0.9\linewidth}}{Pure detection} \\
        \midrule
        $BAC_{sw}$   & Stream-wise balanced accuracy (used for training). Mean of $SENS_{sw}$ and $SPEC_{sw}$\\
        $SENS_{sw}$  & Stream-wise sensitivity. Positive detection rate\\
        $SPEC_{sw}$  & Mean of $SPEC_{pp}$ and $SPEC_{npp}$\\
        $SPEC_{pp}$  & Specificity (negative class accuracy) of blocks (streams) that do not contain a target event, but at times at which in \emph{another} stream, the target event \emph{is} active\\
        $SPEC_{npp}$ & Specificity of blocks that do not contain a target event, at times at which the target event is \emph{inactive in all} streams\\
        \midrule
        $BAC_{tw}$   & Time-wise segregated detection models' balanced accuracy. Mean of $DR_{tw}$ and $SPEC_{tw}$\\
        $DR_{tw}$    & Time-wise segregated detection models' detection rate; aggregated over streams.\\
        $SPEC_{tw}$  & Time-wise segregated detection models' specificity (negative class accuracy); aggregated over streams.\\
        \midrule
        $BAC_{fs}$   & Fullstream models' balanced accuracy. Mean of $DR_{fs}$ and $SPEC_{fs}$\\
        $DR_{fs}$    & Fullstream models' detection rate (positive class accuracy).\\
        $SPEC_{fs}$  & Fullstream models' specificity (negative class accuracy).\\
        \midrule
        \multicolumn{2}{p{0.9\linewidth}}{Localized detection} \\
        \multicolumn{2}{p{0.9\linewidth}}{(all conditioned on target events being active \emph{and} detected)} \\
        \midrule
        $BAPR$     & Best-assignment-possible rate. Proportion of sound event detections in the best-available (azimuth-wise) stream, but in no other stream\\
        $NEP$      & Number of excess positive assignments --- amount of streams with false positive event detections\\
        $AzmErr$   & Mean azimuth error. Averages the azimuth distance of all positive-assigned streams to the correct azimuth\\
        \textit{Placement likelihood} & Depicts the average proportions of event detections in streams depending on their distance to the event's correct azimuth\\
        \bottomrule                             
    \end{tabular}
\end{table}

\begin{figure*}[t]
    \centering
    \subfloat[Time-wise detection]{\includegraphics[height=5.5cm]{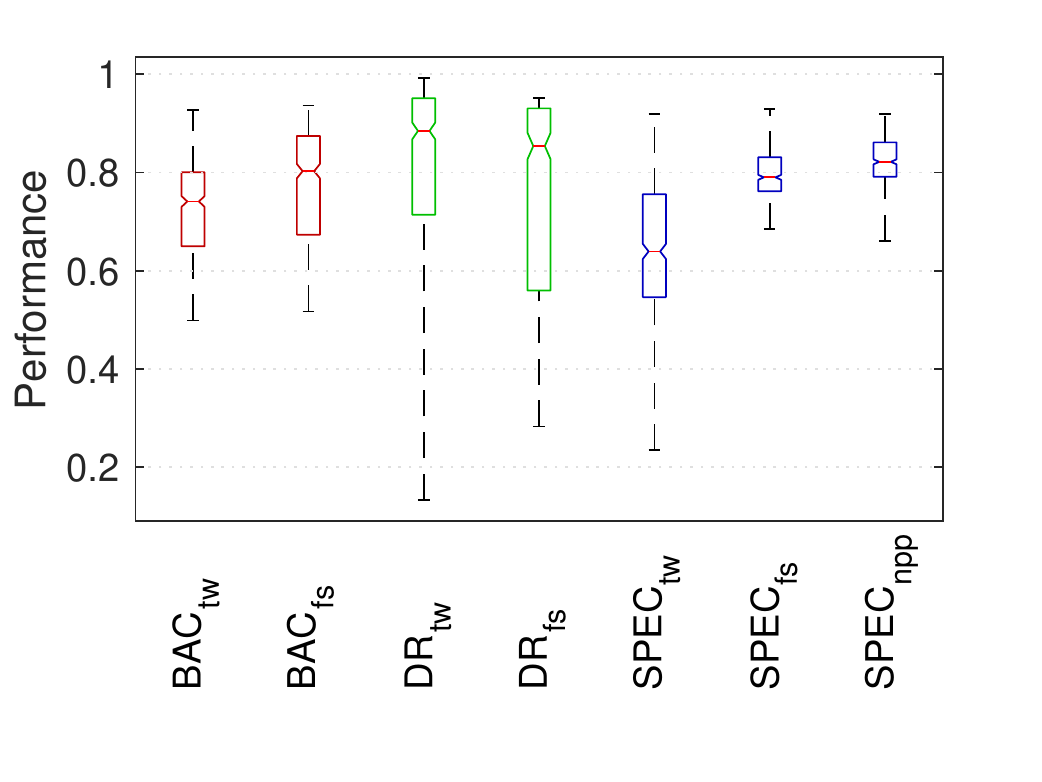}%
        \label{fig:fs_SegId_comparison_grandAverage}}
    \hfil
    \subfloat[Localized detection]{\includegraphics[height=5.5cm]{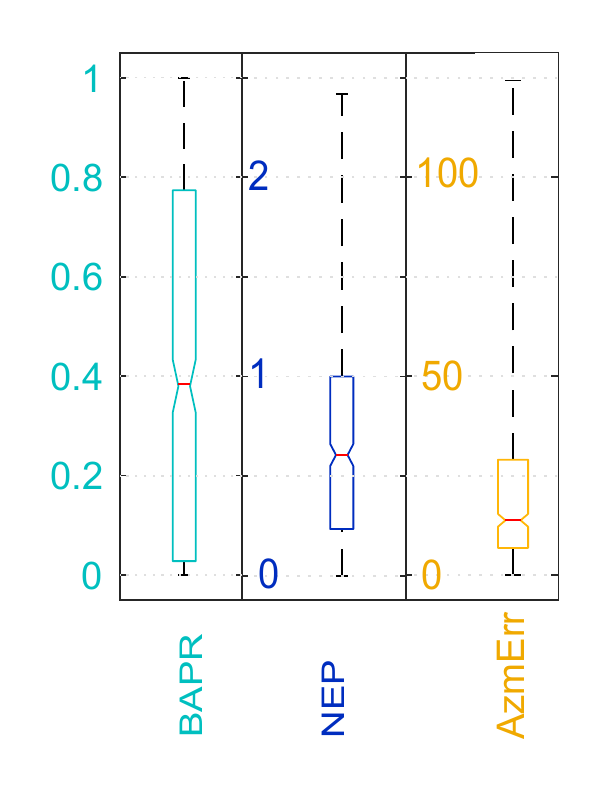}%
        \label{fig:segIdPerf_grandAverage}}
    \hfil
    \subfloat[Placement likelihood]{\includegraphics[height=5.5cm]{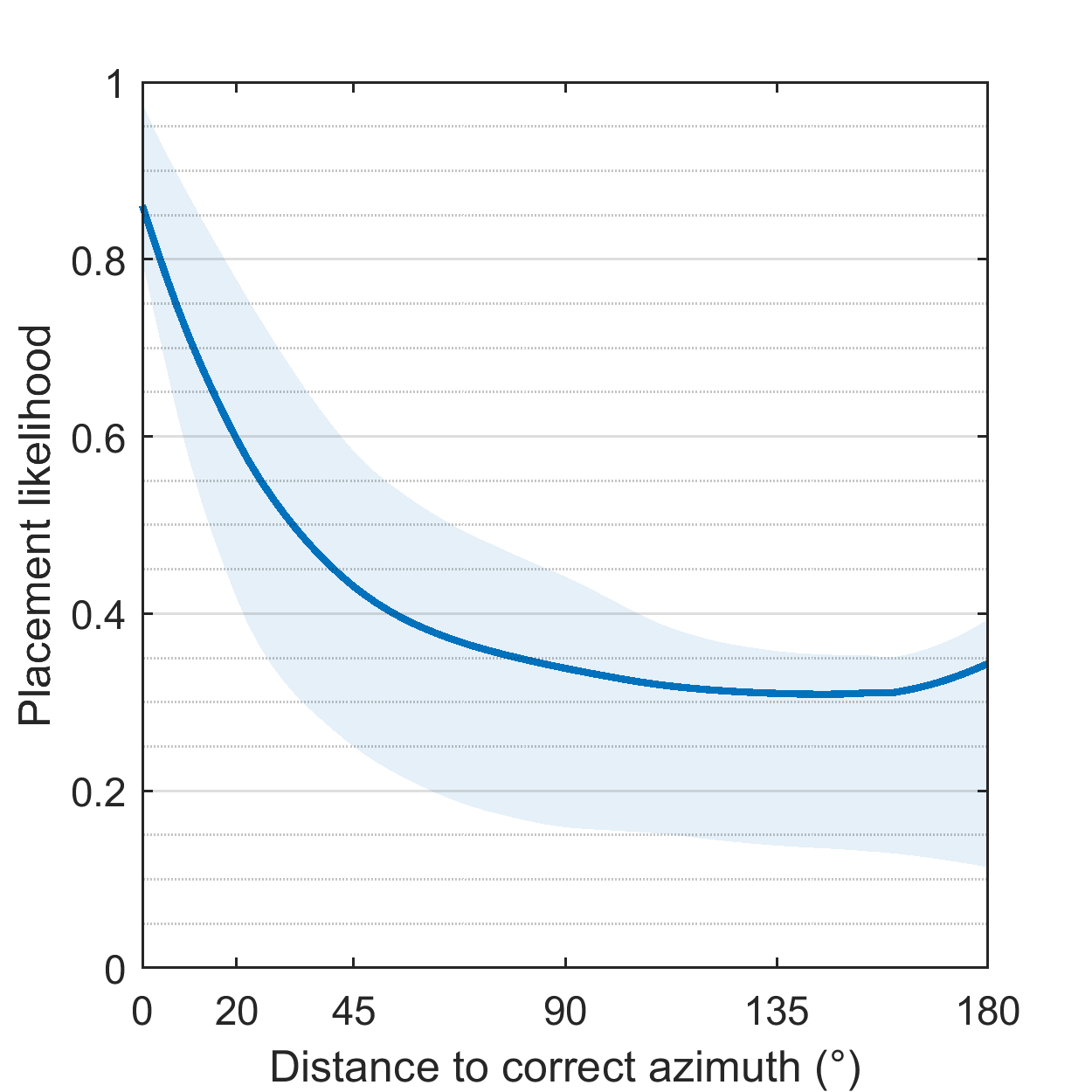}%
        \label{fig:plllh_grandAverage}}
    \caption{Grand average (full test set, all test files, all classes) performances. Time-wise performances (\protect\subref*{fig:fs_SegId_comparison_grandAverage}) are ignorant of location, providing detection performance \emph{aggregated} over streams for segregated detection models, and comparing to fullstream models' detection performance on the full mix. Localized detection performances (\protect\subref*{fig:segIdPerf_grandAverage},\protect\subref*{fig:plllh_grandAverage}) present measures regarding detection in the \emph{correct} stream (that is, associated to the correct location). Box-plots indicate the 25th to 75th percentiles, the median and its \SI{95}{\percent} confidence interval, whiskers depict the complete range of values. The placement likelihood plot (\protect\subref*{fig:plllh_grandAverage}) displays the arithmetic mean and, shaded, the 25th to 75th percentiles. \cref{tab:measures} or \cref{sec:evaluation} provide descriptions of the presented measures.}
    \label{fig:perfs_grandAverage}
\end{figure*}

\subsubsection{Time-wise Detection Evaluation}\label{sec:evaluation_ta}

To evaluate how well the system recognizes sound events irrespective of location and to compare performance to fullstream sound event detection models, we use time-wise measures, namely $BAC_{tw}$, mean of detection rate $DR_{tw}$ \footnote{
Detection rate and sensitivity are the same; the first term is more widely used in SED research.
} and specificity $SPEC_{tw}$.
To obtain these, we \emph{aggregate} the segregated detection models' predictions over streams for each point in time: a positive prediction in any stream produces an aggregate positive prediction. Hence, an aggregate negative prediction is constituted only if all streams are predicted negative. It is obvious that this can lead to an increase of the number of true positives as well as of false positives (shown and discussed in \cref{sec:methodFunctionality,sec:perf_sceneConfig_nsrcs}).

The subindex $tw$ indicates time-wise aggregate segregated detection performance, $fs$ indicates fullstream models' performance.


\subsubsection{Localized Detection Evaluation}\label{sec:evaluation_segId}

To evaluate how well the system assigns detected sound events to the localized streams, we establish four measures that provide understanding of the behavior when a sound event \emph{is present and detected}:
\begin{itemize}
    \item The \emph{placement likelihood} measures the average proportion of streams getting assigned positives, depending on their distance from the sound event's correct azimuth. Ideally, the placement likelihood would be 1 at the correct azimuth, and 0 everywhere else\footnote{Only if the correct azimuth is actually always among the segregated streams, that is, for unperturbed data.}.
    \item The \emph{best-assignment-possible rate (BAPR)} describes how often the system assigns a positive to the stream with associated location \emph{closest} to the true azimuth, and \emph{only} to this stream. For unimpaired source-count and location input (see \cref{sec:methods_perturbation}), the closest stream is always the one with correct azimuth; for perturbed situations, it may well be a stream with azimuth distance greater than \ang{0}.
    \item The \emph{number of excess positive assignments (NEP)} indicates how many streams erroneously got assigned a positive. Ideally, this would be zero.
    \item The \emph{mean azimuth error (AzmErr)} averages the distance of all positive-assigned streams to the correct azimuth.
\end{itemize}

\cref{tab:measures} provides an overview over measures and nomenclature for easy later reference.


\section{Results and Evaluation} \label{sec:results}

In \cref{sec:methodFunctionality}, we demonstrate the general functionality of the proposed method, followed by a discussion about the influence of the main acoustic scene parameters in \cref{sec:sceneDifficulty}. In \cref{sec:locError,sec:nsrcsError}, the impact of estimation errors with respect to locations and number of sources is evaluated.


\subsection{Method Functionality}\label{sec:methodFunctionality}

\begin{table}[t]
    \caption{Generalization: stream-wise performances on full training and test set, averaged over classes and all scenes}
    \label{tab:methodPerformance}
    \centering
    \begin{tabular}{lcc}
        \toprule
        Performance  & test set mean & training set cross-validation mean \\
        \midrule
        $BAC_{sw}$   & $0.777$ & $0.806$\\
        $SENS_{sw}$  & $0.775$ & n.a. \\
        $SPEC_{pp}$  & $0.649$ & n.a. \\
        $SPEC_{npp}$ &$0.837$ & n.a. \\
        \bottomrule
    \end{tabular}
\end{table}

\begin{figure*}[t]
    \centering
    \subfloat[Performances over SNR]{\includegraphics[height=5.2cm]{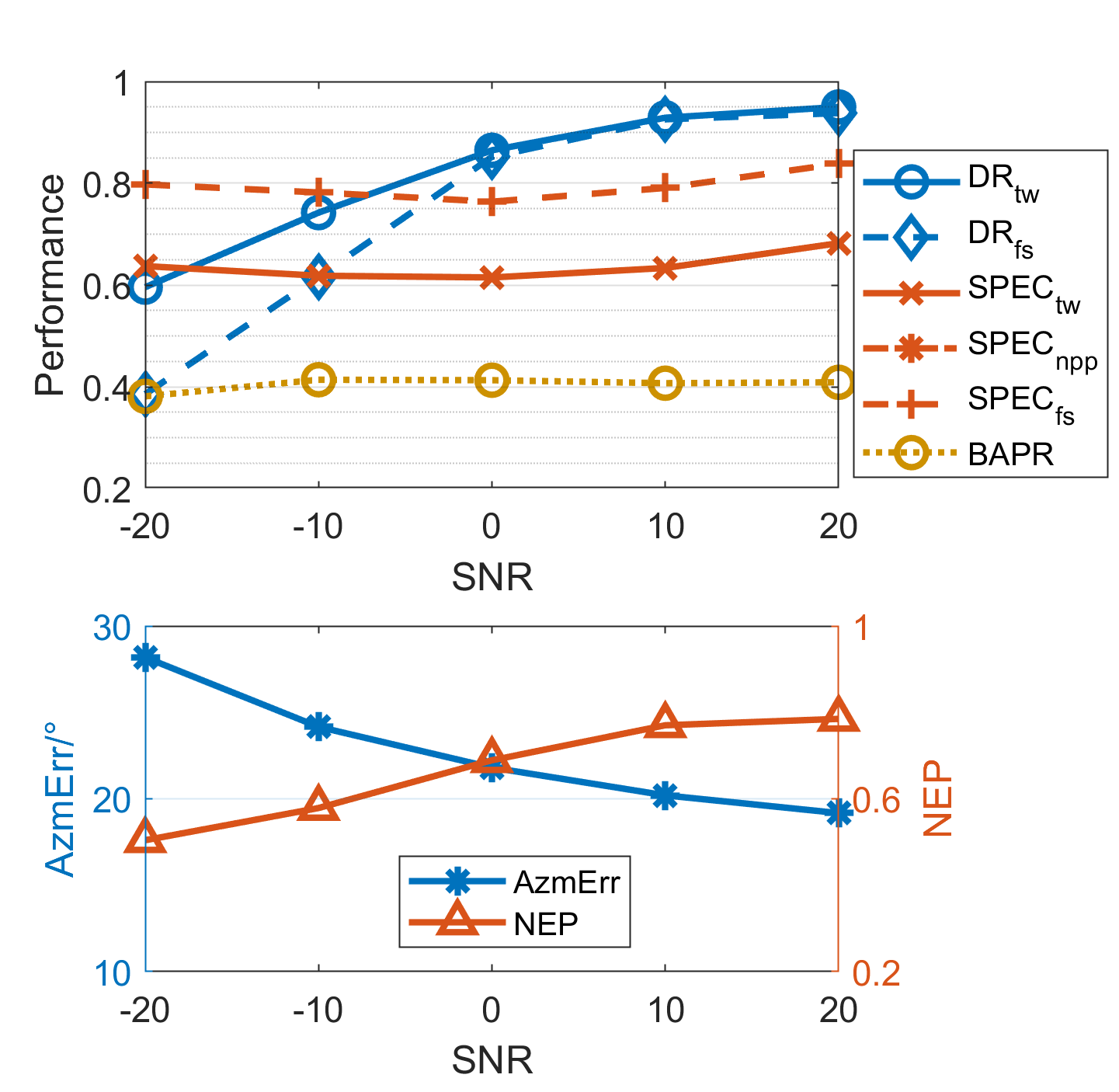}%
        \label{fig:perf_snr}}
    \hfil
    \subfloat[Placement likelihood \newline over SNR]{\includegraphics[height=5.2cm]{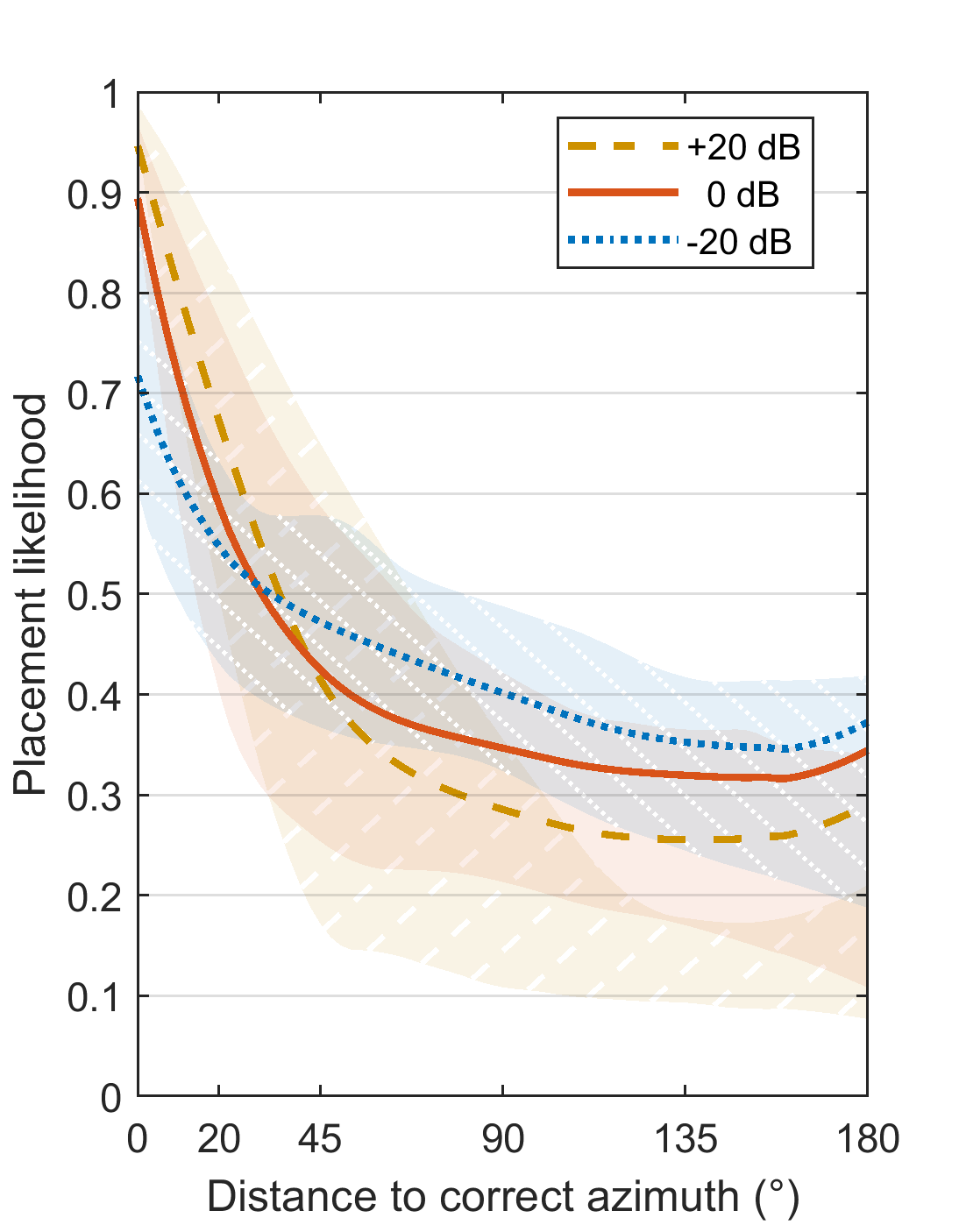}%
        \label{fig:plllh_snr}}
    \hfil
    \subfloat[Performances over \newline number of sources]{\includegraphics[height=5.2cm]{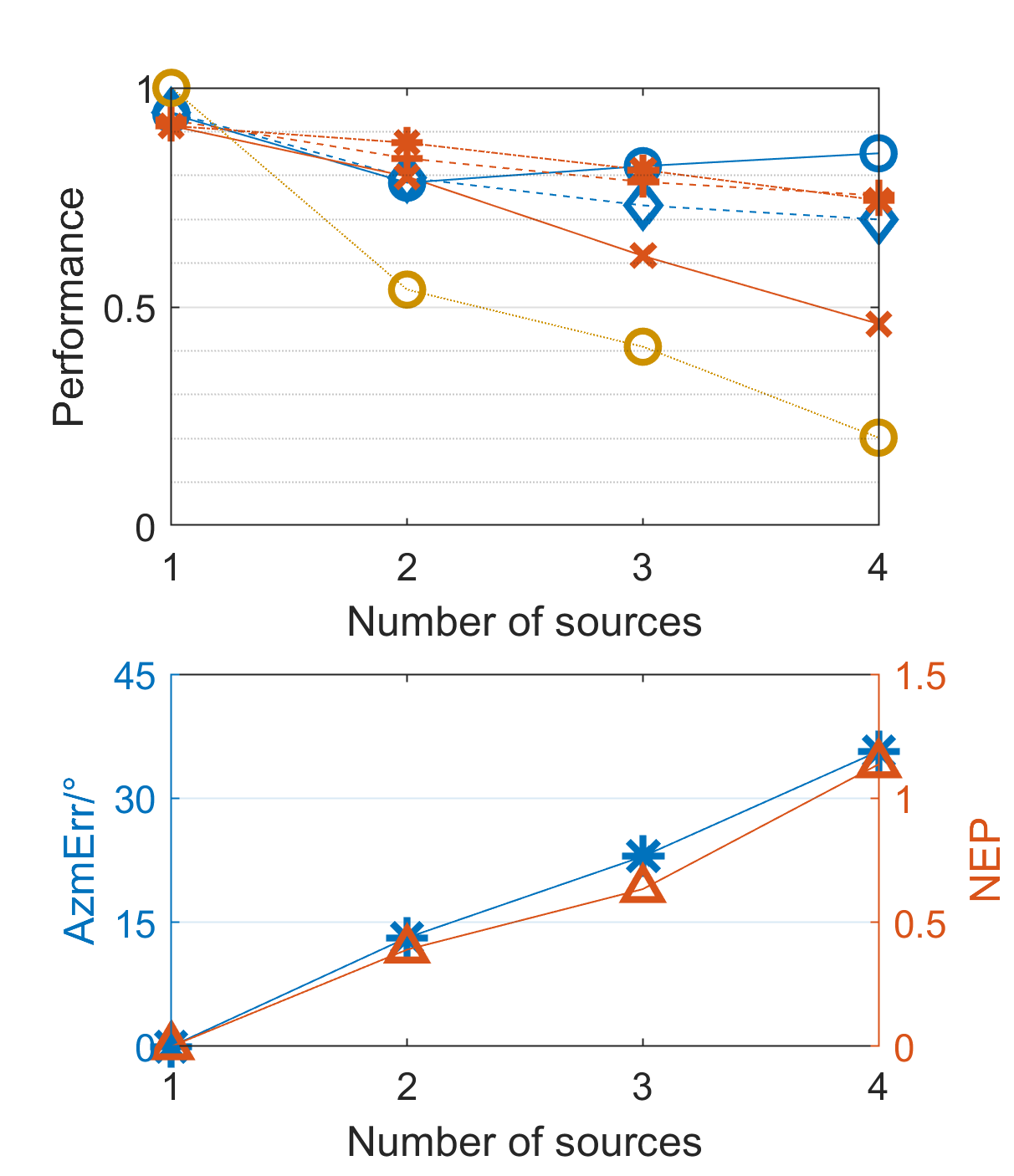}%
        \label{fig:perf_nsrcs}}
    \hfil
    \subfloat[Placement likelihood over \newline number of sources]{\includegraphics[height=5.2cm]{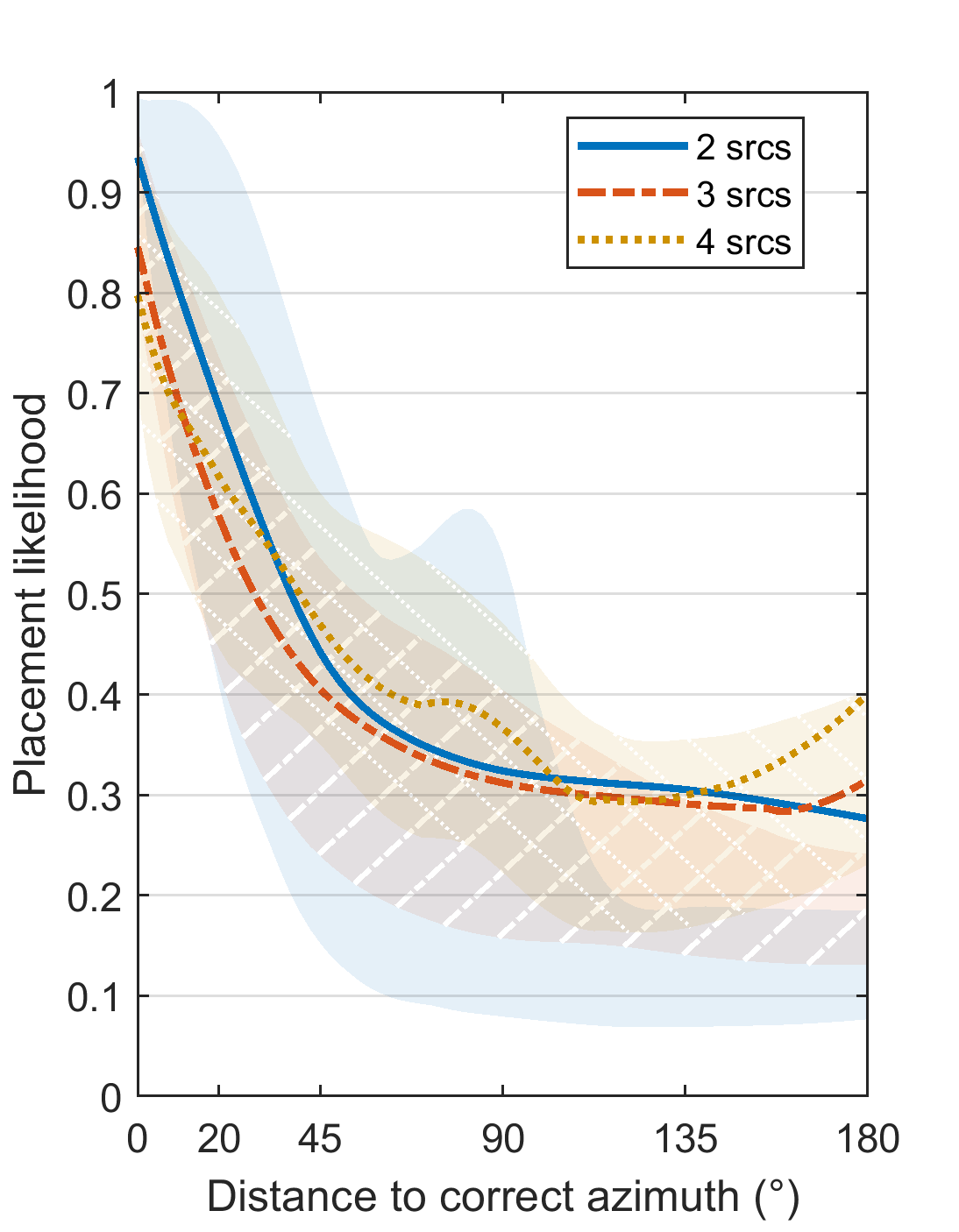}%
        \label{fig:plllh_nsrcs}}
    \caption{Performances depending on SNR (\protect\subref{fig:perf_snr},\protect\subref{fig:plllh_snr}) and number of sources (\protect\subref{fig:perf_nsrcs},\protect\subref{fig:plllh_nsrcs}), averaged over all respective test scenes, all test files, and all classes. Line plots display arithmetic means and, shaded, 25th to 75th percentiles. \cref{tab:measures} or \cref{sec:evaluation} provide descriptions of the presented measures. }
    \label{fig:perfs_nsrcs_snr}
\end{figure*}

Training produces functional models, as \cref{tab:methodPerformance} shows. $BAC_{sw}$  on the test set (averaged such that scenes with 1,2,3,4 sources have equal weight, as in training) is only a bit below training performance and well above chance level. ($SENS$, $SPEC_{pp}$ and $SPEC_{npp}$ constitute $BAC_{sw}$, cf. \cref{sec:methods_training}.) Since different sounds and different scenes are used in the test set compared to the training set, this performance demonstrates successful generalization of the models. 

Disassembling the surrogate performance number $BAC_{sw}$, in \cref{fig:fs_SegId_comparison_grandAverage}, time-wise performances of segregated detection are given and compared to fullstream detection. While being in the same range, the fullstream models do exhibit higher balanced accuracy. This is due to a notably worse specificity of the segregated detection models (median of $0.66$, i.e. one out of three times, without a sound event present, the system actually assigns a positive to one or more streams) for which the better detection rate (median of $0.9$, i.e. nine out of ten times, when a sound event is present, it is also detected) can not make up. This has to be carefully interpreted (see \cref{sec:perf_sceneConfig_nsrcs}), since the additionally depicted underlying \emph{stream-wise} $SPEC_{npp}$ of the segregated detection models is actually even a bit higher than the fullstream models' specificity.

The actual purpose of the segregated detection models is assigning sound events to the correct spatial stream. \cref{fig:segIdPerf_grandAverage,fig:plllh_grandAverage} show different indicators in this regard:
\begin{itemize}
    \item Looking at the placement likelihood\footnote{This graph reads like: if there was a stream located at \ang{20} distance to the true sound source's azimuth, the mean proportion of blocks from this stream getting a positive sound event assignment would be $0.6$.}, a sound event placement is most likely in a stream at the correct azimuth. This likelihood quickly drops with increasing azimuth distance up to around \ang{60}. The ideal system would produce a peak at \ang{0} only, but the graph shows that the method produces assignments more likely to be close to the true azimuth than far from it. 
    \item The best-assignment-possible rate (BAPR) is, in the median, about \SI{40}{\percent}. That is, for the more difficult half of the scenes, between \SI{0}{\percent} and \SI{40}{\percent} of the event assignments are made to the correct stream (and only to it). For the easier half of the scenes, between \SI{40}{\percent} and \SI{100}{\percent} of the assignments are made to the correct stream (and only to it). The wide range indicates that scenes differ a lot in how well they can be segregated into localized streams; which is analyzed and discussed in \cref{sec:sceneDifficulty}. 
    \item The median azimuth error (AzmErr), giving the mean distance between the true sound event's azimuth and the azimuths of its assigned streams, is about \ang{13} and ranges from \ang{0} to \ang{125}. This low average deviation is consistent with the placement likelihood plot, showing that most assignments are done close to the true azimuth.
    \item The median number of excess positive assignments (NEP) is $0.6$. For about \SI{25}{\percent}  of the scenes, only one stream is assigned a positive (which is ideal), but for the larger part of scenes, assignments to more than one stream occur frequently. Looking at the azimuth error, at least these excess assignments usually happen to streams close to the true azimuth.
\end{itemize}



\subsection{Influence of Scene Configuration}\label{sec:sceneDifficulty}

The presented all-scenes grand average results exhibit a very wide range of performance. To investigate the factors influencing performance, three main scene configuration parameters are varied systematically across scenes in the following: the \gls{snr} between target and distractor sources, the number of distractor sources, and the scene mode.

\subsubsection{SNR}\label{sec:perf_sceneConfig_snr}

The performance of the system over different \glspl{snr} is presented in \cref{fig:perf_snr,fig:plllh_snr}. For all \glspl{snr}, the same scene configurations are aggregated, hence the \gls{snr} is the only parameter varied.

Detection rate and specificity show typical behavior --- $DR_{tw}$ dropping with \gls{snr}, $SPEC_{tw}$ remaining mostly constant. Notable are the differences between segregated detection and fullstream models: the offset between specificities remains the same, while the detection rate differs only for difficult \glspl{snr}, where the segregated detection models perform better.

The number of positively assigned streams increases with \gls{snr}, because the stronger target source dominates the time-frequency space and more likely overrides ITDs and ILDs of the weak distractors. On the other side the mean azimuth error decreases with \gls{snr}, implying that --- even though with more of them --- the positive assignments at high \glspl{snr} are closer to the correct azimuth. The placement likelihood graph reflects this as well: up to about \ang{30} azimuth distance, higher \glspl{snr} produce more (percentage-wise) assignments. Above about \ang{30}, it reverses and higher \glspl{snr} of the target produce less assignments.

\subsubsection{Number of sources}\label{sec:perf_sceneConfig_nsrcs}

The performance of the system over different source counts is presented in \cref{fig:perf_nsrcs,fig:plllh_nsrcs}.

A clear negative correlation between number of sources and performance values can be observed, with the notable exception of the detection rate, which counter-intuitively increases slightly from two to four sources. For one and two sources, detection rate and specificity of segregated detection and fullstream models are very similar. The time-wise segregated detection $SPEC_{tw}$ however decreases for higher source counts much more strongly than $SPEC_{fs}$ --- while the \emph{block-wise} $SPEC_{npp}$ shows almost exactly the same behavior as $SPEC_{fs}$. This implies that the model's general ability to classify negatives is actually not lower than that of the fullstream models, and leads us to assume that the reason for both the strong decrease in time-wise specificity as well as for the increase in detection rate is actually the \emph{successful segregation} into streams --- which eases detection of positives, be they true, or be they false, due to sound similarity. In a \emph{mix} of active sources
, any positive (true or false) is less likely detected (this is shown by the detection rate of the fullstream model), but the segregation (to a certain extent) un-mixes. Since all sounds apart from the target class sounds are emitted from all distractor sources, higher number of sources mean higher probability of (false) positive occurrences. Hence, the time-wise aggregation over streams produces an increase in detection rate through true or false positives, and a decrease in specificity through false positives. This is an effect we deem practically unavoidable.
In order to re-balance performance between time-wise detection rate and specificity to increase precision, it may be an option to adjust the training performance measure ($BAC_{sw}$) such that the weight of specificity is increased beyond $0.5$. 

The indicators of localized detection performance, $BAPR$, $AzmErr$ and $NEP$, all show lower performance for higher number of sources. This is to be expected, since more sources imply more overlap in time-frequency-space and thus less distinct segregation masks. In the placement likelihood graph, this is difficult to observe, because the means are very similar, but it can be noted by looking at the shaded indications of the 25th to 75th percentiles.

\begin{figure}[t]
    \centering
    \mbox{\subfloat[Performances over scene modes]{\includegraphics[height=5.2cm]{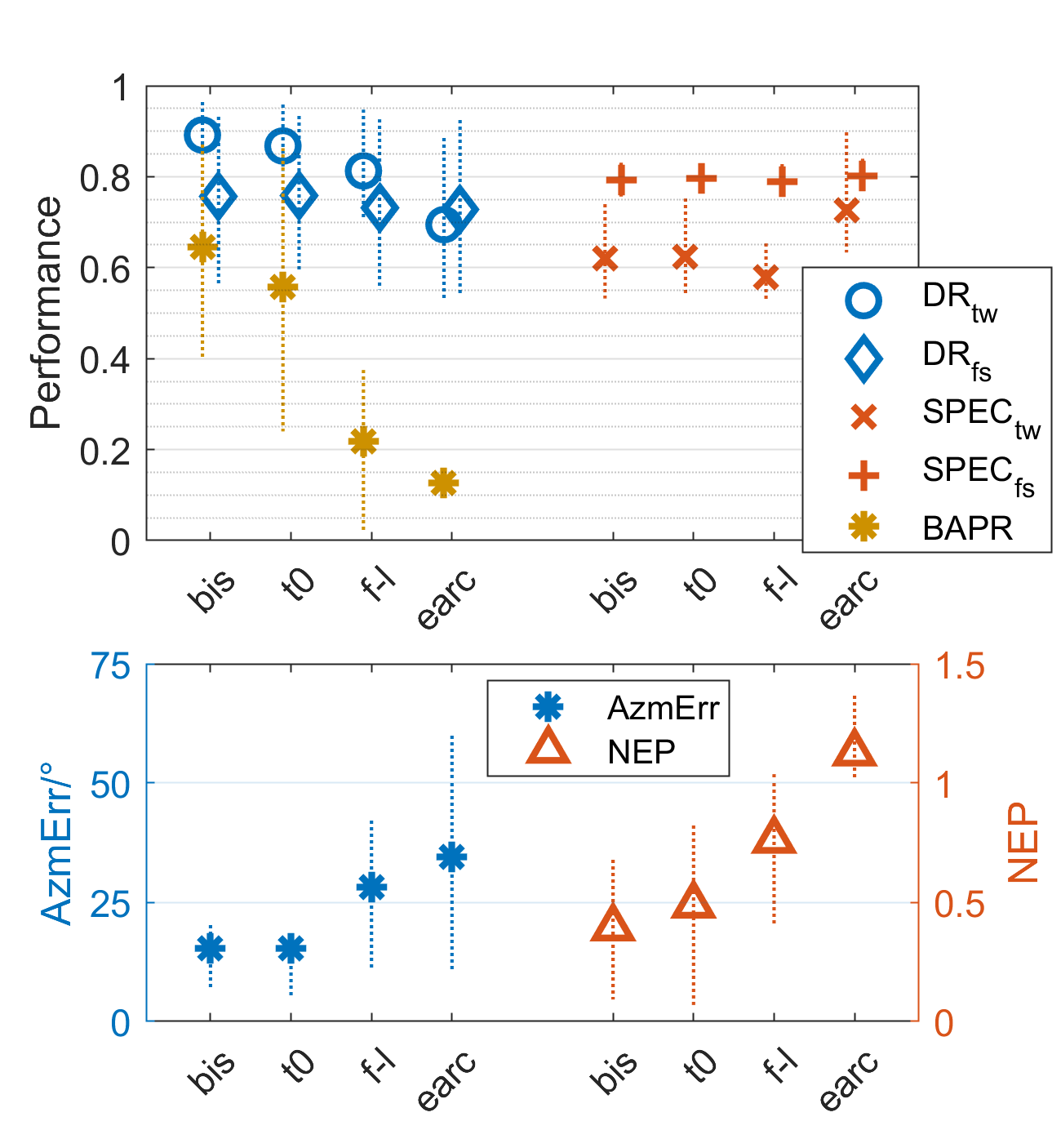}%
            \label{fig:bac_azmModes}}}%
    \hfill
    \mbox{\subfloat[Placement likelihood \newline over scene modes]{\includegraphics[height=5.2cm]{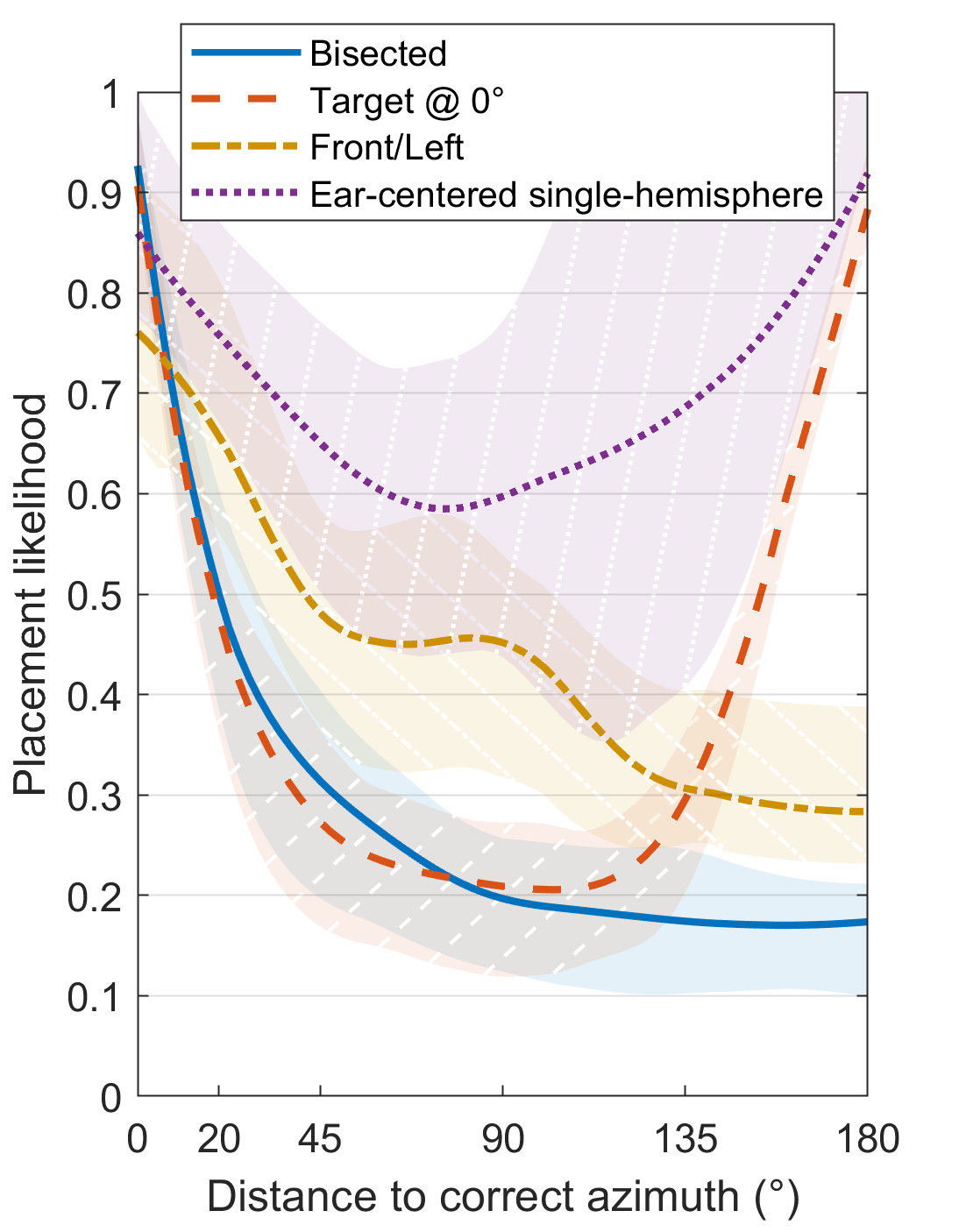}%
            \label{fig:plllh_azmModes}}}%
    \caption{Performances depending on scene mode, averaged over all respective test scenes, all test files, and all classes. Line plots display arithmetic means and, as dotted vertical lines, 25th to 75th percentiles. \cref{tab:measures} or \cref{sec:evaluation} provide descriptions of the presented measures.  }
    \label{fig:perf_azmModes}
\end{figure}

\begin{figure*}[t]
    \centering
    \subfloat[Time-wise performances]{\includegraphics[height=5.5cm]{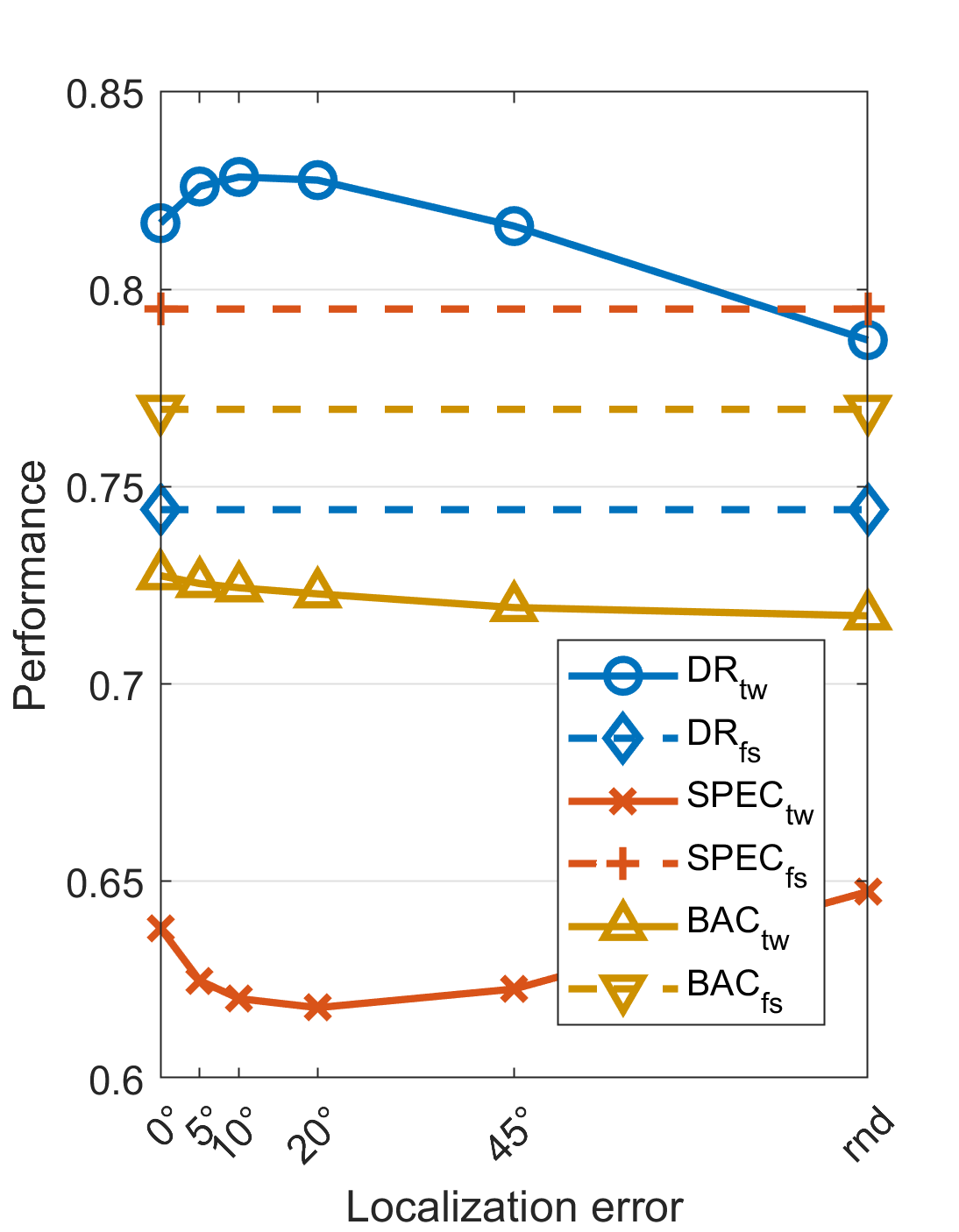}%
        \label{fig:perf_le_grandAverage}}
    \hfil
    \subfloat[BAPR over scene modes]{\includegraphics[height=5.5cm]{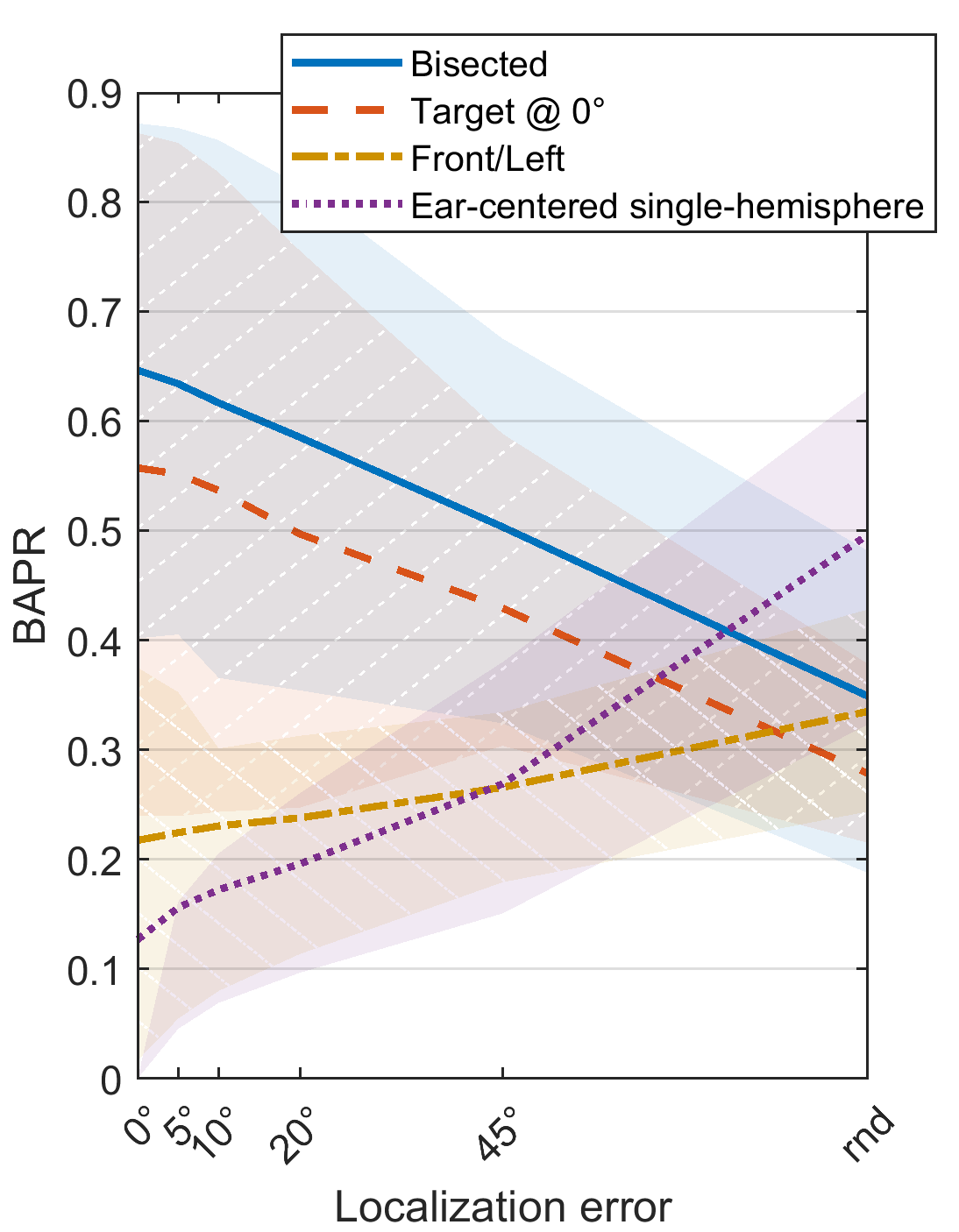}%
        \label{fig:bapr_le_azmModes}}
    \hfil
    \subfloat[Segregated detection performances]{\includegraphics[height=5.5cm]{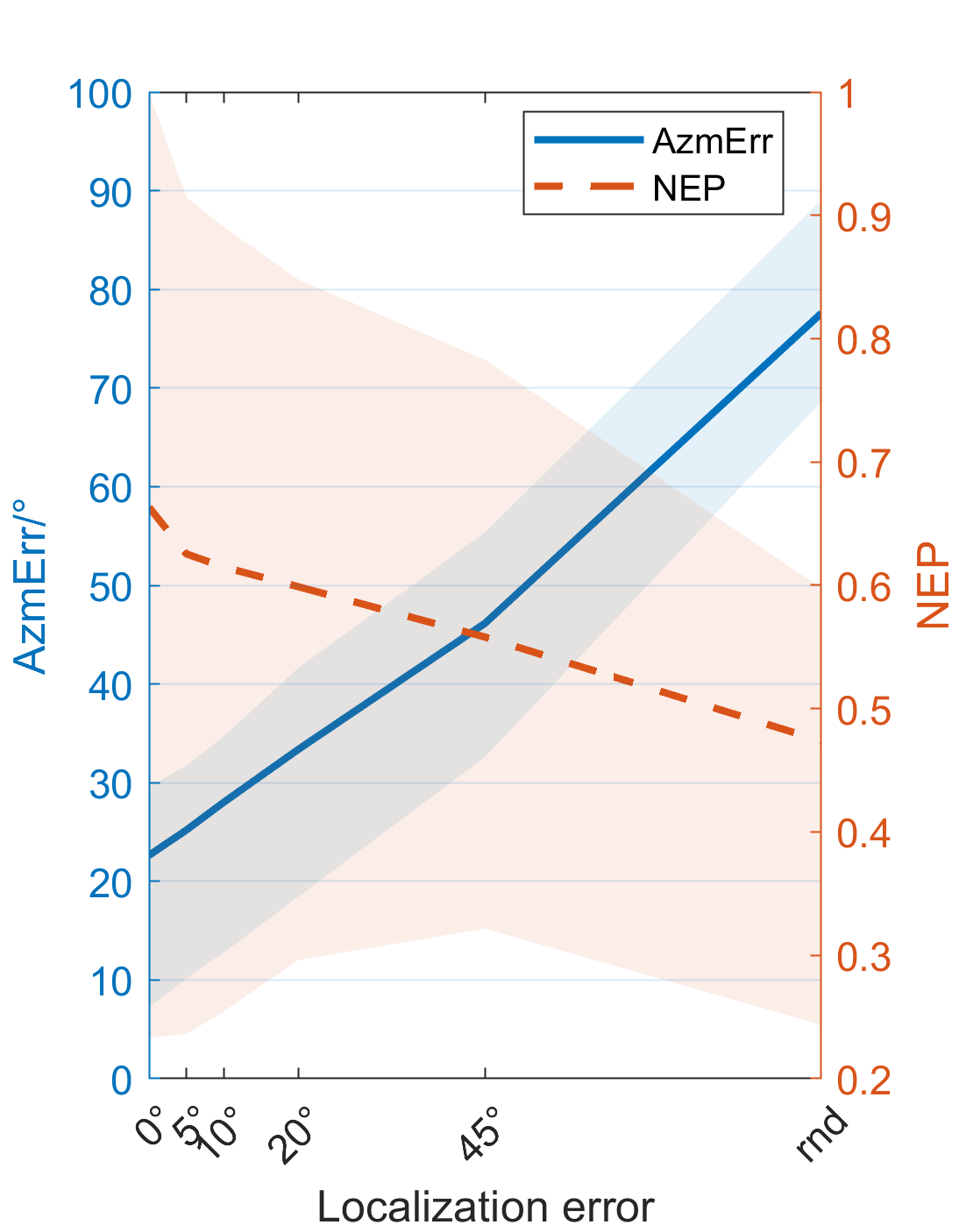}%
        \label{fig:sdPerfs_ae_grandAverage}}
    \hfil
    \subfloat[Placement likelihood]{\includegraphics[height=5.5cm]{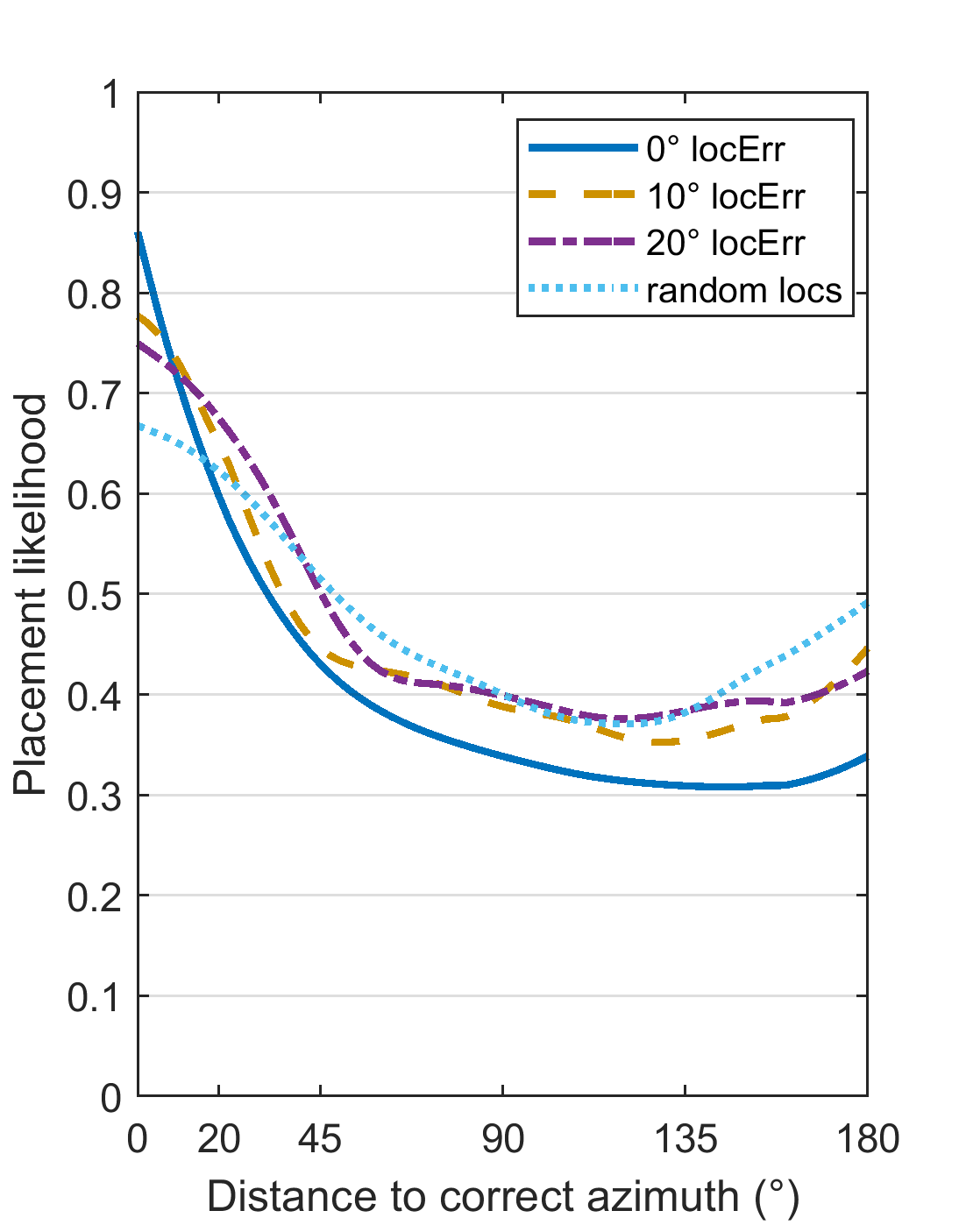}%
        \label{fig:plllh_le_grandAverage}}
    \caption{Grand average (full test set, all test files, all classes) performances depending on strength of perturbation of location information fed into the segregation model. Localization error is given as standard deviation of the Gaussian perturbation added onto true azimuths, ``rnd'' standing for ``random''. Line plots display arithmetic means and, shaded, 25th to 75th percentiles. \cref{tab:measures} or \cref{sec:evaluation} provide descriptions of the presented measures.}
    \label{fig:perf_locError}
\end{figure*}

\subsubsection{Scene mode}\label{sec:perf_sceneConfig_mode}

The performance of the system for the four different scene modes (cf. \cref{sec:scenes}) are presented in \cref{fig:perf_azmModes}. A clear gradation can be observed, with the bisected and target@0 modes performing best, front-left scenes performing worse and ear-centered-single-hemisphere scenes performing by far worst. This holds for all performance indicators apart from specificity. 
Although for the fullstream models the scene mode is far less influential (since the spatial features are not used there), on a much lower scale the same pattern can be noted for the detection rate.

For scenes in bisected or target@0 modes, $BAPR$ is high and $AzmErr$ is low (about \SI{60}{\percent} of all cases with the optimal assignment, and around \ang{15} mean azimuth error), and few excess positives are assigned. 

Particularly the latter increase strongly for the other two modes (negatively correlating $BAPR$) due to the increased occurrence of front-back-confusions. Front-back-confusions emerge because of the (approximate) front-back-symmetry of the head, which leads to similar spatial features for azimuths symmetric to the ear axis \cite{Ma2017}. The employed segregation model for this reason actually disregards differences between front and back at all, in favor of more robust segregation in the frontal hemisphere (cf. \cref{sec:segregation}); any ear-symmetric scene hence must produce equal softmasks and result in the same classification of the symmetric streams.

The placement likelihood graph shows these effects very clearly. The bisected mode shows a curve close to the ideal, while the ear-centered curve demonstrates a severe lack of discrimination between locations for event assignments. Scenes with target at \ang{0} show similar behavior as bisecting ones, but exhibit front-back-confusion approaching \ang{180}.

While \gls{snr} and number of sources are unchangeable attributes of a given scene, the scene mode \emph{is} changeable by head rotation. At least in a scene with sources changing positions slower than the head can turn, it should be possible to notably increase performance by choosing the look direction such that the sources of interest are spread as wide as possible throughout the frontal hemisphere, optimally bisected. This is in accordance with results about dynamically improving localization performance in a binaural robot system \cite{Ma2017}.


\subsection{Detector Performance and Localization Deviation}\label{sec:locError}

The segregation model relies on knowledge of two scene configuration attributes: the number of active sources and the locations (azimuths) of those sources. For the results above, both model inputs have been fed with ground truth. Since a real system likely would not (always) produce correct information about these two aspects, we conducted experiments with systematically perturbed values. 

This section analyzes the influence of perturbation of the location input. The locations fed into the segregation model have an added random Gaussian component (see \cref{sec:methods_perturbation}) with different variances between \ang{5} and \ang{45}. Additionally, tests were performed with completely random azimuth input. For each individual variance, models were tested again on all test scenes and sound files and analyzed as before.

\cref{fig:perf_locError} shows the performances over localization error. Looking at the time-wise detection performance, it is notable that detection rate and specificity behave inversely, and both only change on a small scale (about $\pm0.03$). 

The segregated detection performance indicators show stronger dependency on localization. The best-assignment-possible rate of the four different scene modes (cf. \cref{sec:scenes,sec:perf_sceneConfig_mode}) converge toward similar (low) values with increasing localization error --- particularly the two well-performing modes (bisecting and target@0) decrease strongly. Interestingly, with random localization, the ear-centered scene mode exhibits the best $BAPR$, standing out from the other three modes\footnote{This is because for target sources at \ang{90}, which occur in this scene mode, the probability of any spatial stream with random location getting a similar mask is least. (Highest for \ang{0})}. This can not lead to head turning rules of course, because with random localization, a robot would not know how to position sources at the ear. 

\begin{figure*}[t]
    \centering
    \subfloat[Time-wise performances]{\includegraphics[height=5.5cm]{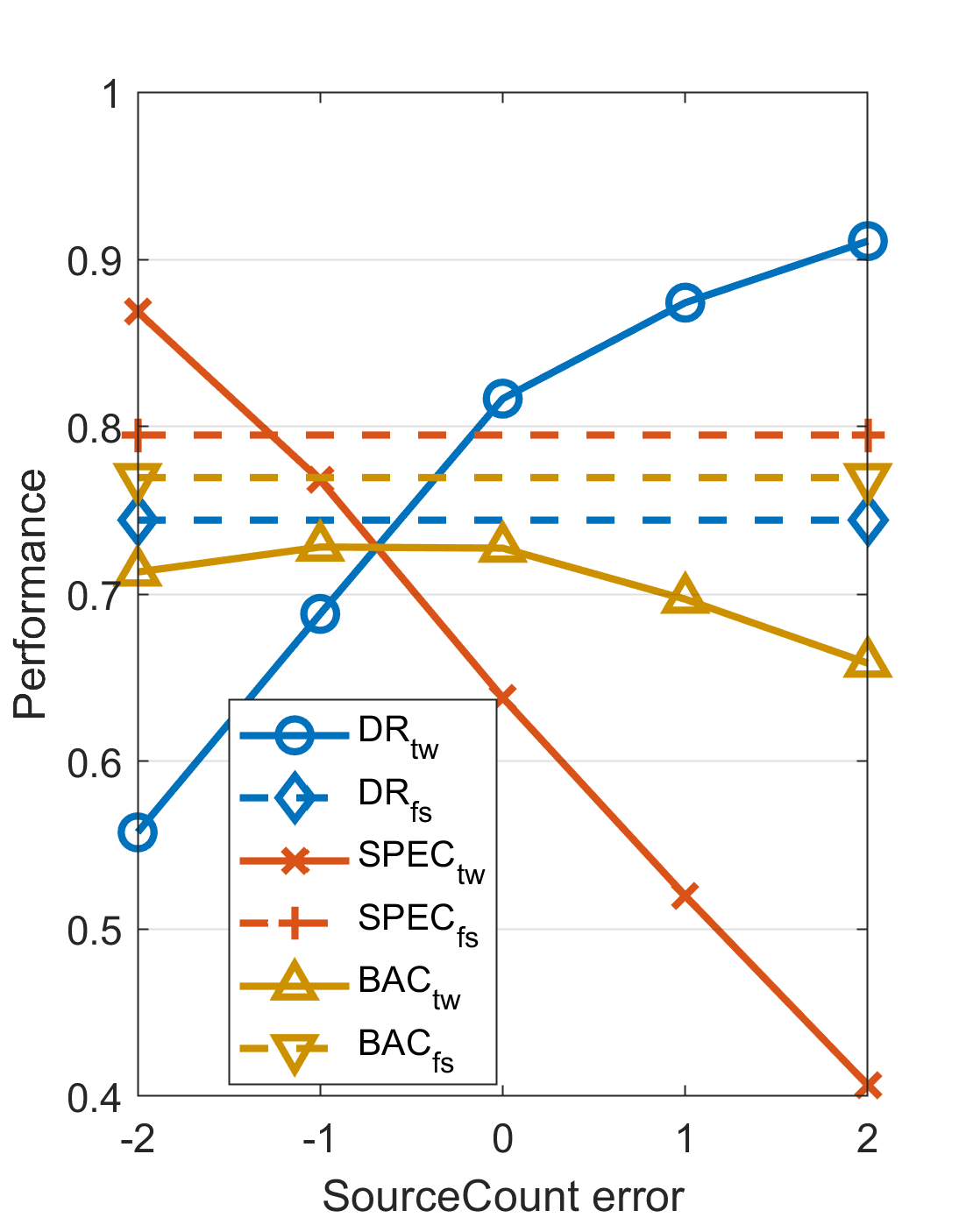}%
        \label{fig:perf_nse_grandAverage}}
    \hfil
    \subfloat[BAPR over scene modes]{\includegraphics[height=5.5cm]{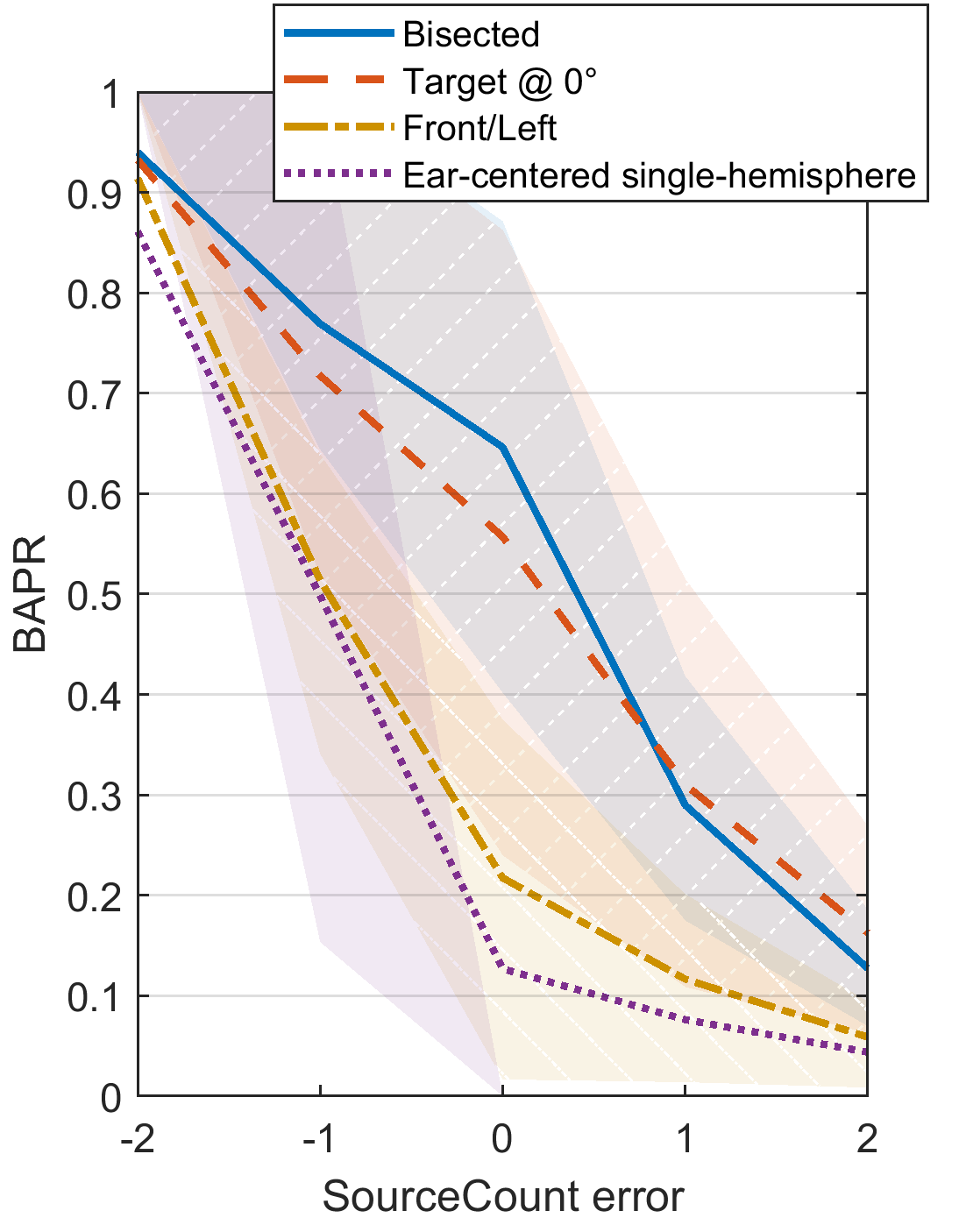}%
        \label{fig:bapr_nse_azmModes}}
    \hfil
    \subfloat[Segregated detection performances]{\includegraphics[height=5.5cm]{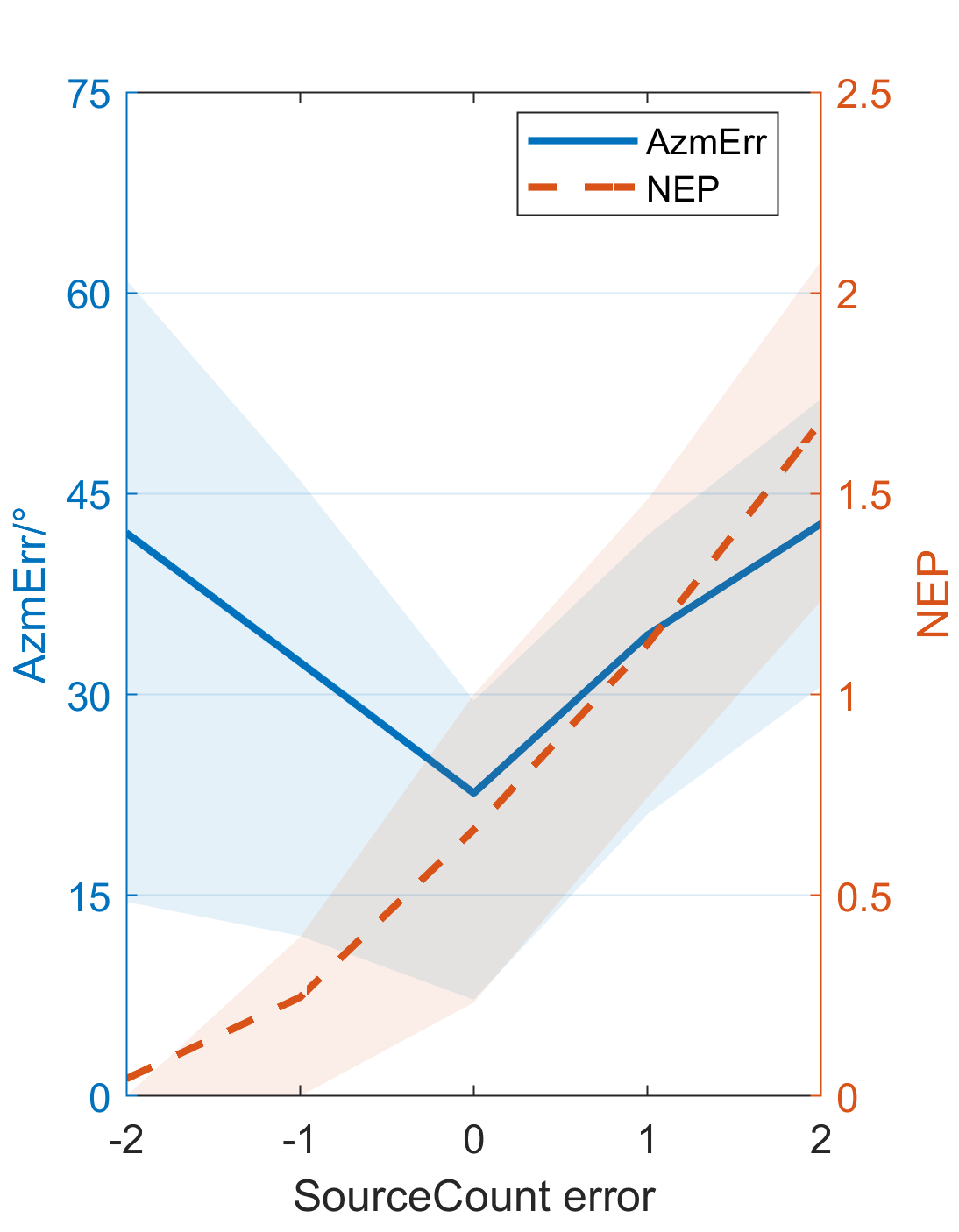}%
        \label{fig:sdPerfs_nse_grandAverage}}
    \hfil
    \subfloat[Placement likelihood]{\includegraphics[height=5.5cm]{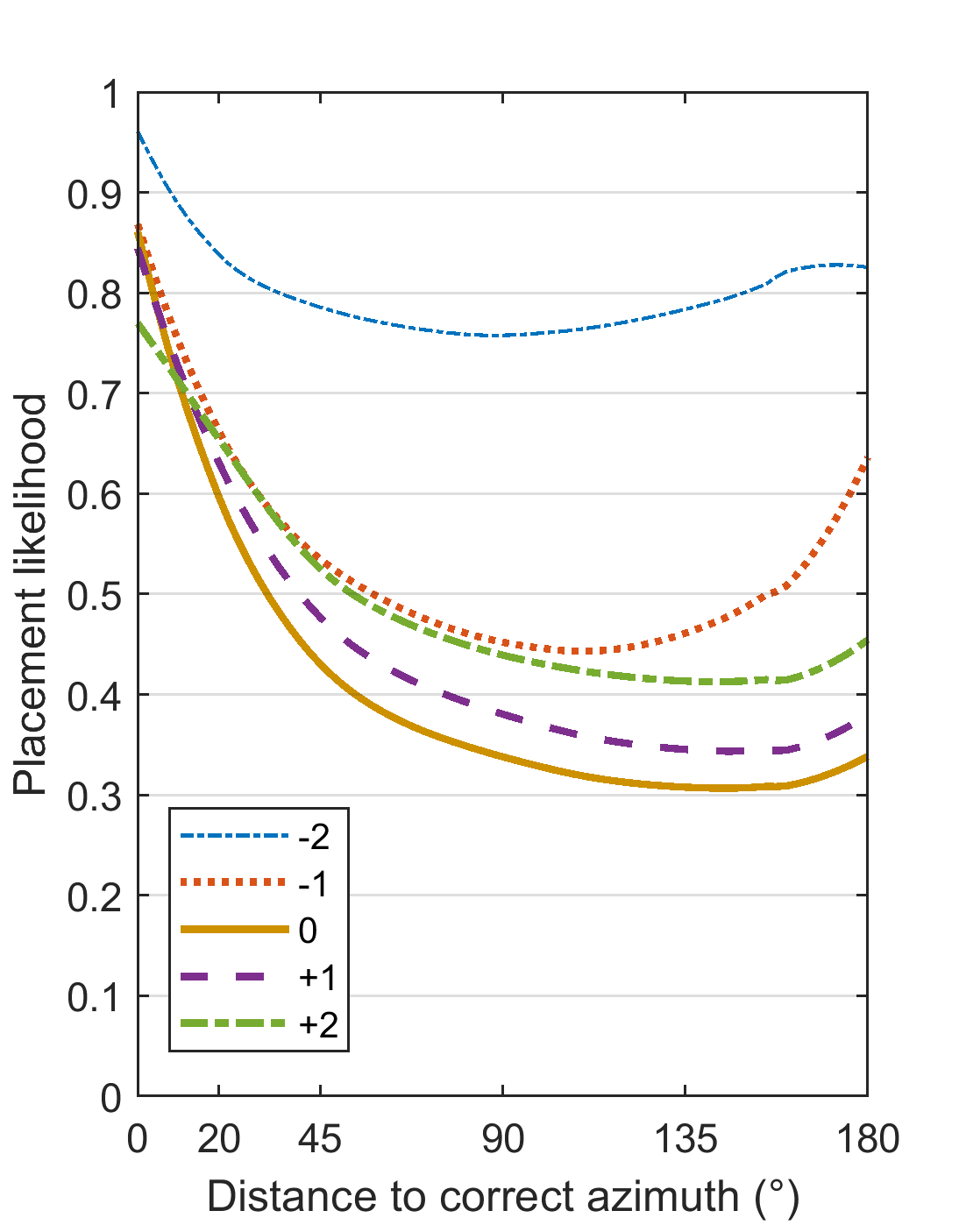}%
        \label{fig:plllh_nse_grandAverage}}
    \caption{Grand average (full test set, all test files, all classes) performances depending on strength of perturbation of source count information fed into the segregation model. Line plots display arithmetic means and, shaded, 25th to 75th percentiles. \cref{tab:measures} or \cref{sec:evaluation} provide descriptions of the presented measures. }
    \label{fig:perf_nsError}
\end{figure*}

Since the performance order of the four modes remains stable up to very high localization errors, the head turning guiding principle deduced in \cref{sec:perf_sceneConfig_mode} stays valid, albeit with lower resulting performance gain.

The straight increase of $ AzmErr $ with localization error is logical --- actually, the localization error does not even fully add to the system-inherent (at \ang{0} localization error) azimuth error of about \ang{22}.

\cite{Ma2017} demonstrated a localization system based on DNN on blocks of \SI{0.5}{\second} with an error of less than \ang{5} for on average \SI{95}{\percent} of all evaluated cases (one to three competing sources, several different reverberant conditions). For a Gaussian distribution, this translates to a sigma of at most \ang{2.5}. The herein tested errors hence should provide upper limits of realistic actual localization with a large margin.


\subsection{Dependence on Number of Sources Estimation}\label{sec:nsrcsError}

After localization error, we analyze the impact of incorrect input of the number of active sources. To this end, we added an error of $ \pm2 $ to the source count and accordingly produced streams by the segregation model. 

\cref{fig:perf_nsError} shows performance over source count error --- it is apparent that deviations from the correct number of streams bear strong performance changes. Time-wise detection rate and specificity show anti-correlated behavior: for underestimation of number of sources, the detection rate degrades heavily, for overestimation of the source count, specificity drops even more.

Azimuth error and placement likelihood show that segregating into the wrong number of streams in both directions leads to worse localized detection performance. The azimuth error rises with any deviation: with too few streams, because the correct stream may be omitted, with too many streams, because segregation becomes more difficult and, as can be noted looking at $ NEP $, because more excess positives are assigned.

The latter is also comprised in the strong decrease of $ BAPR $ for source count overestimation --- any case of excess positive assignment is not a best-possible assignment. The increase of $ BAPR $ for negative source count error is no indicator of somehow better localized detection performance, but a mere logical consequence of the fact that with number of sources underestimation, scenes with two, respectively three, sources become segregated into one stream only, in which case the best-possible assignment is trivial.

Clearly the implication of these results is that the segregated detection system as proposed is dependent on an accurate estimation of number of active sources, but this is true of many systems in computational auditory scene analysis.


\subsection{Effect of Acoustic Perturbation}\label{sec:acousticPerturbation}


\begin{figure*}[t]
    \centering
        \subfloat[Performances over SNR]{\includegraphics[height=5.5cm]{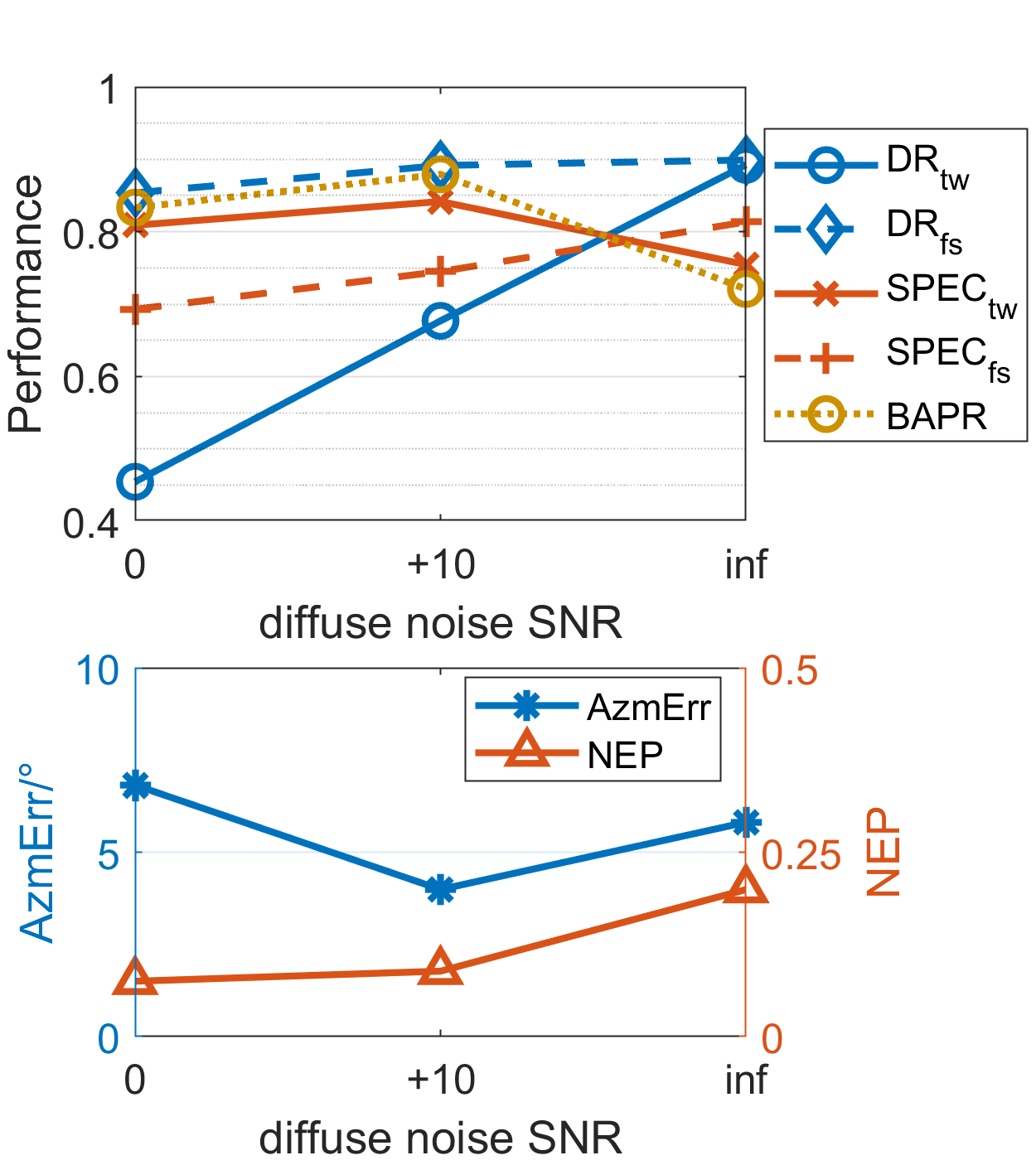}%
            \label{fig:perfs_dn_snr}}
        \hfil
        \subfloat[Placement likelihood]{\includegraphics[height=5.5cm]{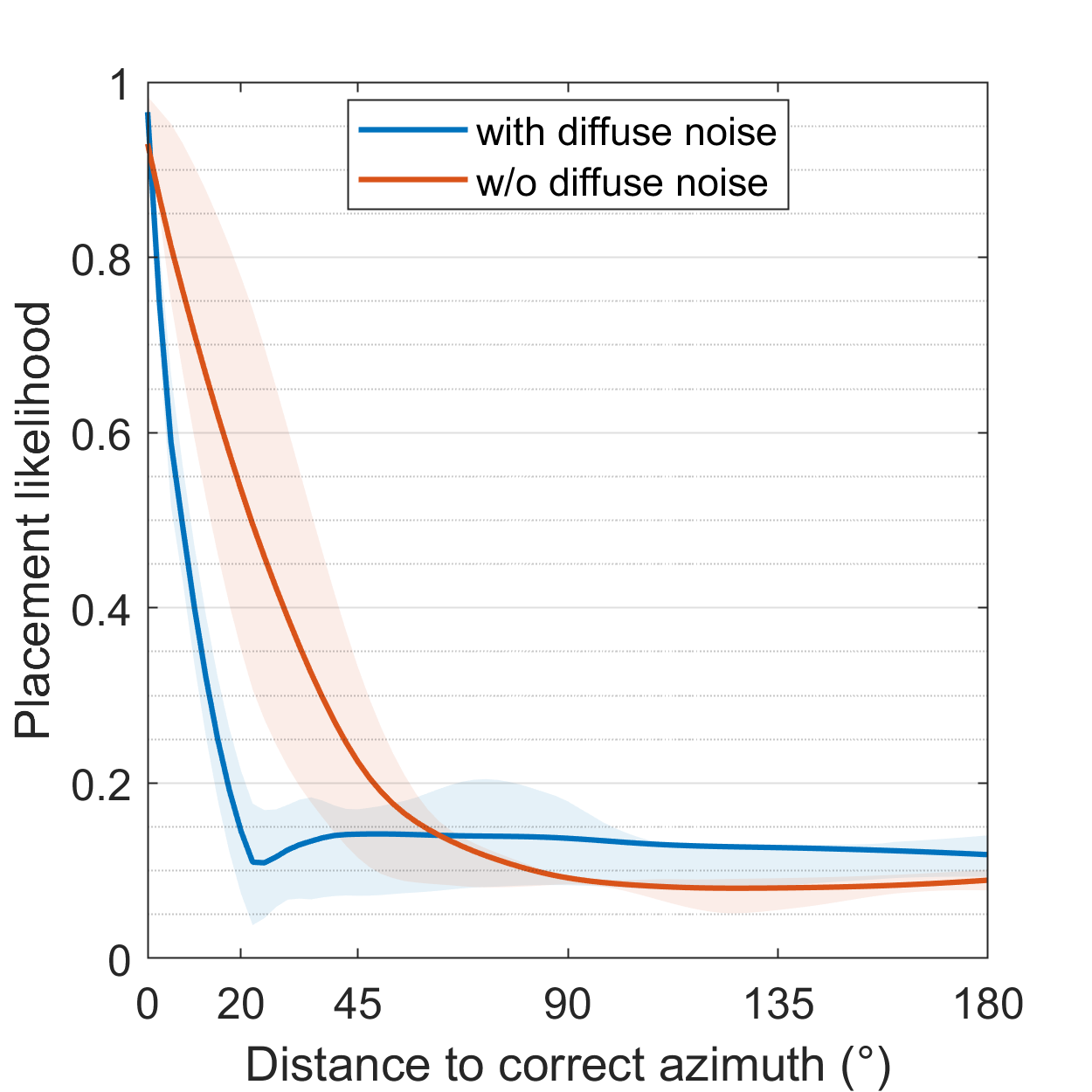}%
            \label{fig:perf_dn_plllh}}
        \hfil
        \subfloat[Performances over scene modes]{\includegraphics[height=5.5cm]{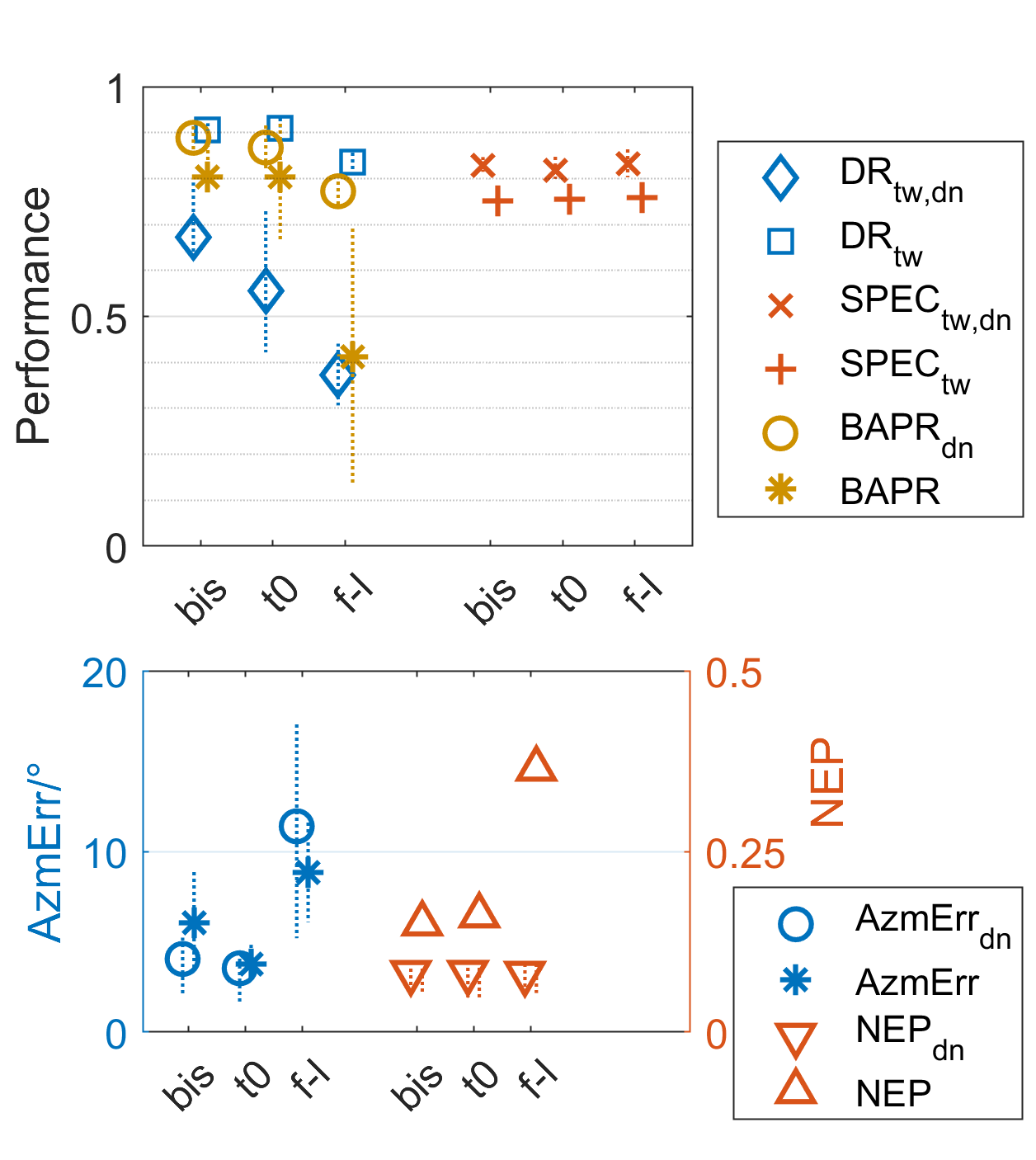}%
            \label{fig:perfs_dn_azmModes}}
        \caption{Performances depending on strength of diffuse white noise, averaged over all respective test scenes, all test files, and all classes. In (\protect\subref*{fig:perfs_dn_snr}), the SNR is expressed from point sources to diffuse noise, $inf$ hence refers to scenes without diffuse noise. In (\protect\subref*{fig:perf_dn_plllh},\protect\subref*{fig:perfs_dn_azmModes}), diffuse noise scenes include both \SI{0}{\decibel} and \SI{+10}{\decibel} configurations. In (\protect\subref*{fig:perfs_dn_azmModes}), the subindex $dn$ refers to scenes with diffuse noise. Line plots display arithmetic means and, shaded or as dotted vertical lines, 25th to 75th percentiles. \cref{tab:measures} or \cref{sec:evaluation} provide descriptions of the presented measures. }
        \label{fig:perf_dn}
\end{figure*}

While the analyses presented do include very challenging situations, there are acoustic settings that adversely affect the perceived spatial definedness of point sources, which were untouched in above experiments. Specifically, \emph{diffuse noise} and \emph{reverberation} are frequently encountered in real acoustic environments.

\subsubsection{Diffuse noise}\label{sec:perf_diffuseNoise}

Strictly speaking, all acoustic sources are spatially localized, that is, emitted by (more or less) point sources. However, perceptually, there can also be a share of spatially diffuse noise, produced by a huge number of individual point sources that are perceived together. Usually these individual point sources will be caused by a common process, like for example rain (individual drops falling all around) or traffic noise in the middle of a very busy place.

To be able to estimate the effect, which such diffuse noise would have on the proposed segregated detection system, we added diffuse noise to a subset\footnote{using the full test set was infeasible} of the test scenes. Specifically, we added diffuse \emph{white} noise at SNRs (point sources to diffuse noise) of \SI{+10}{\decibel} and \SI{0}{\decibel} to scenes with two point sources\footnote{the two point sources having an SNR of \SI{0}{\decibel} to each other} in bisected, target@0, and front-left modes. The diffuse source was simulated by $360$ independent white noise point sources placed in \ang{1}-steps around the head. The models were then tested on these scenes and the results were compared to those without diffuse noise on the same subset of scenes.

\cref{fig:perf_dn} displays performances of the segregated detection comparing the scenes with and without diffuse noise.
In \cref{fig:perfs_dn_snr}, performances are presented over the SNR of the point sources to the diffuse noise; \cref{fig:perf_dn_plllh} shows the placement likelihood of scenes with and without diffuse noise. \cref{fig:perfs_dn_azmModes} plots the performances over scene modes. 
Three relevant observations are to be taken from these two figures:
\begin{itemize}
    \item The diffuse noise impairs the detection rate. The fullstream model is showing a mild effect on pure sound event detection. However, the segregated detection models show a strong degradation with increasing diffuse noise. Since the difference between the two model types are the applied segregation masks, this implies that strong diffuse noise can corrupt these masks in many cases (for an SNR of \SI{0}{\decibel}, about $DR_{fs} - DR_{tw} = \,$\SI{40}{\percent}).
    \item On the other hand, the stable values of $BAPR$, $AzmErr$, $NEP$ and the similar placement likelihood curves show that for the cases in which the sound events \emph{are} detected, the method produces equally good assignments to the spatial streams as in the scenes without diffuse noise.
    \item The spatial scene arrangement has the same impact as without diffuse noise, but even stronger.
\end{itemize}
While the latter two observations are reassuring, the effect on the segregated detection rate is severe: this must be attributed to the influence of the diffuse noise on the ITDs and ILDs (which are the features used by the segregation model), making them noisy as well (which is caused by the diffuseness rather than by the type of the noise signal). The problem of these features is that the noise can override the information generated by the spatially distinct point sources. The usage of artificial \emph{white} noise of course complicates the task for the models compared to noise from real-life sources, which rarely will be white and occupying all time-frequency bins.
In the discussion, we will elaborate on possible solutions to be able to deal even with strong diffuse noise.

\subsubsection{Reverberation}\label{sec:perf_reverb}

The analyses presented were conducted with anechoic acoustic scenes. In open-space or natural environments like forests, this can be a realistic situation; however, in rooms, in cities close to buildings, etc, usually (moderate) reverberation will be present, and of course there exist highly reverberant environments like churches. Reverberation mixes acoustic signals in time and space and particularly perturbs spatial cues, and hence poses challenges for any kind of model relying on spatial information of sound. 

This also held true in a few additional experiments with reverberant scenes we conducted. For these experiments, we simulated acoustic scenes in the ``Auditorium 3'' of Technische Universit\"at Berlin, a mid-sized lecture hall, through the binaural room impulse responses published at \cite{wierstorf2016audi3}. Since we expected significant changes of spatial features and produced segregation masks, we retrained both the segregation model and the detection models with Auditorium 3-scenes. The trained models were then tested on (of course different) scenes also placed in this room.

\cref{tab:brirPerformance} compares the performance of segregated detection for scenes in the reverberant Auditorium 3 with scenes of the same spatial arrangement and SNRs, but in anechoic free-field conditions showing average performance over the respective scenes.

The results clearly show the effect of the reverberation in this room: the localized detection measures $BAPR$, $AzmErr$ and $BAC_{sw}$ are significantly worse than in the equivalent anechoic scenes. The detection rate of the retrained models increases a bit, but at the price of much lower specificity. The system still works ($BAC_{sw}$ significantly above $0.5$, still \SI{49}{\percent} of all detected sound events with the best-possible assignment), but definitely not as well.

Of course reverberated sound \emph{is} more difficult to localize. However, as in the case of strong diffuse noise above, the performance degradation should neither be attributed simply to the reverberation nor to the method itself, but rather to the employed simple components of the system, particularly the segregation model and the ITD and ILD features it uses. In the discussion below, we suggest ways to improve the individual components, and argue that such improvements would make the proposed method much more robust to reverberation.

\begin{table}[t]
    \centering
        \caption{Comparison of reverberant and anechoic performances}
        \label{tab:brirPerformance}
        \begin{tabular}{lcc}
            \toprule
            Performance  & Auditorium3 & Anechoic \\
            \midrule
            $BAC_{sw}$   & $0.66$      & $0.80$\\
            $DR_{tw}$    & $0.77$      & $0.74$ \\
            $SPEC_{tw}$  & $0.68$      & $0.86$ \\
            $BAPR$       & $0.49$      & $0.77$ \\
            $AzmErr$     & \ang{21.8}  & \ang{7.7} \\
            \bottomrule
        \end{tabular}
\end{table}


\section{Discussion}\label{sec:discussion}


\emph{Joint} sound event localization and detection is a new field (at least with respect to machines performing it \footnote{but, as far as the authors of this work understand from the literature known to them, it is also not yet fully understood with humans \cite{bizley2013and}}), unsurprising, considering that even SED does not have a very long history. Publications tackling it are hard to find, and then often do not describe true SELD systems, because identities and locations are predicted side by side \cite{Lopatka2016,butko2011two}, not solving the problem of how to associate them in presence of co-occurring sound sources.

In this work, an approach for binding the prediction of the two modalities has been developed and analyzed. The proposed method combines spatial masking in time-frequency-space with sound event detection on the segregated streams, enabling formation of coherent auditory objects with location and sound event type associated. It could be shown that SED and SSL can be joined efficiently with this method in a modular system, and that robust performance can be achieved through multi-conditional training. 

Systematic analyses with respect to different acoustic conditions have been presented, evaluating the effects of the number of simultaneously active sources, their respective energy ratios, and particularly regarding the influence of spatial source distribution around the head, which is something that has not been done before in this context. Strong impact of the true source locations on the system's performance has been found; optimally, sources are separated by the nose. In a binaural robotic system, source locations are subject to the head orientation, hence we propose to turn the head such that favorable relative source positions can be achieved.

\subsubsection*{Methodology}

Along with the proposed system, the general problem of joint SELD has been introduced and discussed. As part of the contribution of this work, measures for the quantification and qualification of localized sound event detection success have been developed and presented. The solutions obtained include two metrics that similarly have been developed independently in \cite{adavanne2018sound}: what is called here mean azimuth error (AzmErr), is called there ``DOAerror'', and what is called here ``number of excess positives'' (NEP) and time-wise specificity and detection rate, is summarized in their work through the ``frame recall''. The ``placement likelihood'' presented here is unique and adds more fine-grained information about the system's behavior in a comprehensible way.

The performance demonstrated by our system was achieved operating on short segments (\SI{500}{\milli\second}), in contrast to the other works introduced following approach \emph{3}, which operate on the full events (\cite{Grobler2017,Chakraborty2014,May2012binaural}). Likely, using longer segments would increase performance on two ends: first, with longer blocks, sound events should become more identifiable (depending on the type). Second, longer blocks should make streams more segregable, because the strength of superposition varies strongly over time and hence longer blocks will include more frames with the individual sources standing out.

The most similar system was published in \cite{May2012binaural}, performing speech detection and localization bound through spatial segregation in time-frequency space, plus speaker identification, using GMMs. 
Several differences facilitating the task and decreasing generality of their results are worth mentioning: (a) full events were processed there, (b) their system possessed ground truth about the number of active \emph{speech} sources; making the speech detection effectively only determine which streams are the most likely speech streams, (c) only speech was a target type, which in our experiments consistently showed to be the most detectable sound type of all classes present in NIGENS.
The most notable difference however lies in the training paradigm; the system presented in \cite{May2012binaural} was trained on clean (single-source) data, and spatial masks were applied only during testing, together with a missing data approach (\cite{Cooke200126robust}).
This is a clear contrast to the system proposed here, for which robustness to missing data is \emph{learned} through training multi-conditionally in polyphonic scenes with spatial masks already applied. 
The results presented in \cite{May2012binaural} show a strong degradation of the final speaker identification for \gls{snr} even above \SI{0}{\decibel}, while the performances of the herein developed system only degrade very slightly for this \gls{snr} range.

The choice of training performance measure has a crucial impact on system performance in any machine learning model. Compared to employing standard $BAC$, that does not distinguish between $N_{pp}$ and $N_{npp}$ samples and hence would result in models largely unable to assign events to only the correct stream, the proposed $BAC_{sw}$ has shown to produce functioning models. However, there may as well be more suitable measures; in particular, $BAC_{sw}$ does not impose cost on azimuth distance of positive assignments\footnote{Other than discriminating between correct or incorrect stream}. This would be an interesting point for further research.

Because of the significantly lower time-wise specificity, it may be beneficial to combine segregated detection models for localized detection together with fullstream detection models for actual sound event presence detection. The latter would prime or even trigger the application of segregated detection. \cite{adavanne2018sound} basically do the same to reduce false positives.

\subsubsection*{Scalability}
The system could easily be extended to unseen sound types by training additional SED models; since binary one-vs-all SED models (including the ``general'' sounds as counter-examples in the ``all'' part) are used, already-trained SED models would be unaffected. The segregation model does not need to be retrained, since it never was trained on the sound events in the first place, but only on white noise.

The method demonstrates good performance in a lot of situations using only two channels --- as stated in the introduction, this work originated from the {\scshape Two!Ears} project, which aimed for being comparable to humans. But additionally, the two-channel constraint may regularly also be indicated from a purely technical standpoint in, e.g., robotics systems with limited hardware capabilities due to space, energy consumption or computational constraints.

However, the proposed method does not necessarily need to operate binaurally and could easily scale up. Particularly, a segregation model using spatial information from more
channels should be able to produce more accurate masks; obviously, this way it will be easier to often/always be in the ``bisected'' spatial arrangement with its advantageous properties (see \cref{sec:perf_sceneConfig_mode}).

\subsection*{Model Power}

Given the simplicity of the employed segregation and detection models (linear regression), the good results for big ranges of scene configurations confirm our assumption that the modular structure of this approach is favorable for ``low-power'' modeling, which allowed us to perform a wide analysis with a multitude of experiments. Generally, the aim of this work was to perform and provide extensive tests and qualitative analyses of how to fundamentally tackle sound event localization and detection, instead of demonstrating the highest performance possible. Definitely, using a potent, up-to-date model class, with the available computational resources a lot of the presented work would not have been possible to conduct.

However, as seen with the diffuse noise and under reverberation, the segregation model as used here is limited in power. Two main problems are hindering more effective spatial masking under difficult acoustic conditions:
\begin{itemize}
    \item The use of ITDs and ILDs as spatial features, which both only represent spatial information about the dominant source of each T-F-bin, and hence are vulnerable to noise. Instead, it seems advisable to use interaural cross-correlation features (which are very high-dimensional though), which include spatial information about all sources. This is also the representation used for example with the localization model dealing with reverberant environments described in \cite{Ma2017}.
    \item Looking at human auditory scene analysis (ASA), the approach of segregating sounds by treating individual time-frequency bins independently certainly falls short. As \cite{bregman1993complex} reasons, it is more likely that spatial cues are used together with rules about sound regularities. It is reasonable to assume that a segregation model that takes into account \emph{spatio-spectro-temporal context} should improve the produced masks significantly, particularly in noisy situations. 
\end{itemize}
Both points call for a more complex model type: DNNs with their representation- and context-modeling capabilities would be a natural choice. We have no doubt that such a model could improve the spatial stream-formation strongly; but the development of a model like this would be a work of its own --- one that would be nicely placed and benchmarked in the testbed of the herein developed framework.

Certainly, also the detection model class (logistic regression) is a limit to performance in our investigation. Pilot tests with nonlinear SVM (RBF kernels) and random forests did not improve performance --- which led us to assume that basic nonlinear information in the data is already extracted in the feature creation. However, it is reasonable to assume that more complex models able to model and integrate information over time, like \glspl{cnn} or \glspl{rnn}, would show higher performance. 
Such Powerful SED models able to integrate information over time should also be more capable to deal with noisy segregation masks that can emerge from reverberant scenes or scenes with strong diffuse noise.

\section{Conclusion}\label{sec:conclusion}

We have suggested and evaluated a method to annotate sound scenes with joint sound event type and location information in a binaural system. The proposed method combines spatial segregation in time-frequency-space with sound event detection on the segregated streams. Through this combination, the system helps forming coherent auditory objects with associated attributes location and type. 
Our focus was a general demonstration of the concept, development of methodology, and particularly depth in analysis with respect to scene arrangement.

To achieve robustness with respect to varying scene conditions, we propose training multi-conditionally as in \cite{Trowitzsch2017Robust} and, importantly, to perform segregation already in training and not only in testing. 
The presented analysis demonstrates that this approach can produce localized sound type information under a broad range of conditions and could be one core component of a binaural scene analysis system. The localized detection performance depends particularly on the number of active sources in the scene, and on their spatial distribution.
By turning the head such that the sources of interest are in the frontal hemisphere (and at best bisected by the nose), the system's performance in many situations can be increased strongly.

Proper estimation of the number of active spatial streams is a precondition of this approach; a deviation of more than one leads to very strong performance degradation. Localization error, on the other hand, does not influence segregated detection as heavily. Even with large errors, the system does not break down, but propagates the input error under mild impairment of assignment precision to the output.

A diverse set of test scenes for thorough study of the behavior and conditions of good performance was defined along with several performance indicators to enable capturing different qualitative aspects of the joint behavior of the combined models. They can serve, together with the code, as testbed and benchmarks for alike systems with different components and other approaches to the problem, and of course particularly for spatial segregation and sound event detection models. All algorithms and tools for training, testing and evaluation, the sound data and the trained models themselves are provided as public domain (\cite{amlttp30,trowitzsch2019nigens,segIdPDrepo}).

\appendices

\section*{Acknowledgment}

This work was part of the {\normalfont \scshape Two!Ears} project funded by the European Union's Seventh Framework Programme for research, technological development and demonstration under grant agreement no 618075.

Thank you to two anonymous reviewers for their feedback which was very valuable in improving the manuscript.

\ifCLASSOPTIONcaptionsoff
  \newpage
\fi



\bibliographystyle{IEEEtran}
\bibliography{references}
\begin{IEEEbiographynophoto}{Ivo Trowitzsch}
received his Diplom degree in Computer Engineering from the Technische Universit\"at Berlin, Germany, in 2011. He is a graduate research assistant at the Neural Information Processing Group at Technische Universit\"at Berlin. His research interests currently focus on the application of machine learning in computational auditory scene analysis and in general audio processing.
\end{IEEEbiographynophoto}
\begin{IEEEbiographynophoto}{Christopher Schymura}
(S'13---M'17) received the Dipl.-Ing. degree in electrical engineering and information technology from Ruhr University Bochum, Germany, in 2013. Since 2013, he has been working as a research assistant at the Cognitive Signal Processing group, Institute of Communication Acoustics at Ruhr University Bochum. His research interests include machine learning and probabilistic modeling for multimodal signal processing and robotics.
\end{IEEEbiographynophoto}
%
%
%
%
\begin{IEEEbiographynophoto}{Dorothea Kolossa}
Dorothea Kolossa received the Dipl.-Ing. degree in computer engineering and the Dr.-Ing. degree in electrical engineering from Technische Universit{\"a}t Berlin, Germany, in 1999 and 2007, respectively. From 1999 until 2000, she worked on control systems design at DaimlerChrysler Research and Technology, Hennigsdorf, Germany. She was employed as a research assistant at Technische Universit{\"a}t Berlin from 2000 until 2004, and as a senior researcher from 2004 until 2010, also staying at UC Berkeley's Parlab as visiting faculty in 2009/2010. Since 2010, she has been working in a faculty position at the Institute of Communication Acoustics, Ruhr-Universit{\"a}t Bochum, Germany, where currently, she is heading the Cognitive Signal Processing group as an associate professor. Her research interests include speech recognition in adverse environments and robust classification methods for communication and technical diagnostics. She has authored and co-authored more than 100 scientific papers and book chapters. Dr. Kolossa is a member of the Council of the EAA TC on Audio Signal Processing, and of the IEEE Audio and Acoustic Signal Processing Technical Committee.

\end{IEEEbiographynophoto}
\begin{IEEEbiographynophoto}{Klaus Obermayer}
Klaus Obermayer received his Diplom degree in physics in 1987 from the
University of Stuttgart, Germany, and the Dr. rer. nat. degree in 1992 from
the Department of Physics, Technical University of Munich, Germany.
From 1992 and 1993 he was a postdoctoral fellow at the Rockefeller
University, New York, and the Salk Institute for Biological Studies, La Jolla, USA. From 1994 to 1995 he was member of the Technische Fakult{\"a}t, University of Bielefeld, Germany. He became associate professor in 1995 and full professor in 2001 at the Department of Electrical Engineering and Computer Science of the Berlin University of Technology, Germany. He is head of the Neural Information Processing Group and member of the steering committee of the Bernstein Center for Computational Neuroscience Berlin. He was member of the governing board of the International Neural Network
Society from 2004 - 2012 and was Vice-President of the Organisation for Computational Neuroscience from 2008-2011. From 1999-2003 he was one of the directors of the European Advanced Course of Computational Neuroscience. His current areas of research are computational neuroscience, artificial neural networks and machine learning, and the analysis of neural data. He co-authored more than 250 scientific publications.
\end{IEEEbiographynophoto}



\end{document}